\documentclass[num-refs]{wiley-article}
\usepackage{graphicx}
\usepackage[space]{grffile}
\usepackage{latexsym}
\usepackage{textcomp}
\usepackage{longtable}
\usepackage{tabulary}
\usepackage{booktabs,array,multirow}
\usepackage{amsfonts,amsmath,amssymb}
\usepackage{natbib}
\usepackage{url}
\usepackage{hyperref}
\usepackage{etoolbox}
\usepackage{subfig}
\usepackage{float}
\usepackage{subfloat}
\usepackage{longtable}
\usepackage{xcolor,pifont}

\makeatother
\newif\iflatexml\latexmlfalse

\AtBeginDocument{\DeclareGraphicsExtensions{.pdf,.PDF,.eps,.EPS,.png,.PNG,.tif,.TIF,.jpg,.JPG,.jpeg,.JPEG}}

\usepackage[utf8]{inputenc}
\usepackage[english]{babel}

\usepackage{siunitx}

\newcommand*\colourcheck[1]{%
  \expandafter\newcommand\csname #1check\endcsname{\textcolor{#1}{\ding{51}}}%
}
\colourcheck{black}
\newcommand*{\field}[1]{\mathbb{#1}}%
\newcommand*{\m}[1]{\mathbf{#1}}%
\newcommand*{\bb}[1]{\boldsymbol{#1}}%
\newcommand\at[2]{\left.#1\right|_{#2}}
\makeatletter

\iflatexml

\else

\paperfield{Spatial-temporal Models}
\corremail{osafu.austin.in@gmail.com}
\presentadd{Department of Statistics, Institute of Mathematical Science and Computing, University of Sao Paulo, Sao Carlos 13566-590, Brazil}
\fundinginfo{Francisco Louzada acknowledges support from the Brazilian institutions FAPESP and CNPq. Osafu Augustine Egbon acknowledges support from the Brazilian institution CAPES.}

\papertype{Original Article}

\title{Spatial Statistical Models: an overview under the Bayesian Approach}

\author[1]{Francisco Louzada}

\author[1]{Diego C. Nascimento}

\author[1]{\text{Francisco Louzada \hspace{1.0cm}  Diego C. Nascimento \hspace{1.0cm}  Osafu Augustine Egbon}\\
		Institute of Mathematical Science and Computing, University of Sao Paulo, Sao Carlos 13566-590, Brazil}

\runningauthor{Osafu Egbon}
\begin{document}

\maketitle
\selectlanguage{english}
\begin{abstract}
Spatial documentation is exponentially increasing given the availability of \emph{Big IoT Data,} enabled by the devices miniaturization and data storage capacity. Bayesian spatial statistics is a useful statistical tool to determine the dependence structure and hidden patterns over space through prior knowledge and data likelihood. Nevertheless, this modeling class is not well explored as the classification and regression machine learning models given their simplicity and often weak (data) independence supposition. In this manner, this systematic review aimed to unravel the main models presented in the literature in the past 20 years, identify gaps, and research opportunities.  Elements such as random fields, spatial domains, prior specification, covariance function, and numerical approximations were discussed. This work explored the two subclasses of spatial smoothing \emph{global} and \emph{local}.
\end{abstract}

\section*{Introduction}
Digital transformation technologies generate massive amounts of data in the past decades, labeling these concepts as \emph{Big Data}, in which data storage grows exponentially and requires an advanced analytic tool to explore and answer research questions. The technical advancement created open doors to model complex phenomenons such as spatial trends and heterogeneity across space and time, such as applying spatial methods on daily observed data in smart cities and urban informatics to identify and predict high risk regions\cite{tao2013interdisciplinary, roche2016geographic}. In this manner, the \emph{Internet of Thinks} (IoT) is reshaping daily tasks, in which miniaturization devices are also placing location labels much easier than before \cite{marjani2017big}. Spatial dependencies have long been identified as a component that could hinder model precision and increase bias. Subsequent effort to account for such error created a research line in spatial statistics. 

    Observation oriented across space is an essential feature for the IoT data, which influences data prediction and analysis. The applications vary in complexity, and it is frequently applied in risk surfaces detection, healthcare, agriculture, urban planning, economics, and rarely applied in smart appliances that learn based on location. The complex structure is accommodated in a flexible class of models related to the observed data and the spatial dependencies. The frequentist (classical) and the Bayesian analytical methods have been used to analyze IoT data. However, the Bayesian method is a better choice because it owes its ability to accommodate information from different sources. In the Bayesian framework, elucidated questions are answered in the estimation procedure by combining multiples sources of information, such as previous knowledge (prior) and the acquired information in the data (likelihood) \cite{razafimandimby2017bayesian}.

In the neuroscience field, neurorehabilitation has growing and technological improvements made given the neuro-navigation, allowing personalizing the definition of transcranial accuracy~\cite{harquel2017automatized,meincke2016automated}. Recently, there is an increasing interest in using spatial models on the meta-analysis of brain imagery to locate hot spot regions of consistent activation on the brain for diagnostic and treatment \cite{derado2013,kang2011}. Another exemplification is in epidemiology, which takes advantage of patterns across geographical space to identify the areas of potentially elevated risk and create disease maps to quantify underlying risk surface \cite{kang2016making}. Moreover, the medical literature provides detailed motivation and descriptions of spatial smoothing methods by explaining the concepts, defining the technical terms, and demonstrating various visualizing spatial models.

The basic idea of the Bayesian spatial statistics is the extension of the generalized linear model, including a spatial component that accounts for spatial dependencies across a spatial domain. 
The component is assigned a spatial prior, usually multivariate, which accounts for spatial correlation across a region (not necessarily delimited). Afterward, the parameter estimates are smoothed across the spatial domain with a specified resolution to identify the hot spot region and provide intuition on the chain of events. Moreover, the approach is different for frequently used spatial econometric models that treat the spatial dependencies as a global correlation parameter across a spatial domain. In the heart of every spatial model have a  correlation matrix that quantifies the dependencies of the spatial component and determines the complete distribution of the spatial prior. The correlation matrix, proportional to the weight matrix, has an enormous impact on spatial smoothing, and the challenge usually faced with authors is the choice of its specification.
Additionally, the model complexity can be owed to the structure of the weight matrix. A highly dense weight matrix implies high correlated spatial field and a more complex model. Besides the general specification of the diagonal element of the weight matrix set to zero, over the past 20 years, there has not been a concrete documented standard on the specification of the off-diagonal elements. The most frequently used weight matrix is the binary first neighborhood structure, which assigns 1 or 0 depending on whether the spatial locations are immediate neighbors or not. 

\subsection*{Objectives of this review} 
A Bayesian spatial statistic aims to quantify the spatial pattern and provide insight into the process generating the pattern.
 Given the technological advances and precisely the storage capacity, data georeferenced acquisition is commonly present in the nowadays domain sets. The Bayesian method is typically used to analyze these sets and to identify the spatial pattern. Consequently, Spatial Statistics have received considerable attention in recent years, and numerous spatial models have been proposed and applied in diverse research fields. The systematic review of these models has received little attention and specifically been conducted in epidemiology \citep{aswi2019bayesian,duncan2017spatial,wah2020systematic}. 
\\This systematic review focused on the progressive development and the content analysis of the Bayesian spatial models and to bridge the discontinuities in the literature. It aims to provide an overview and the basic knowledge of the concepts, improvements for the last 20 years, and identify the key research directions and areas of opportunities in the Bayesian spatial methodology. 

\subsection*{Outline}
R. A. Fisher has long identified the implication of spatial dependence in statistical analysis \citep{Fisher1935}. He introduced blocking in a complete randomized design to mitigate the error induced by spatial dependencies. For several years, there have not been many changes from the basic idea of Fisher characterization of spatial dependencies such that close locations are assumed to have a similar trend. In this manner, we aimed to enlighten this vital topic towards the popularization and development of the spatial modeling field.

This systematic review is structured in three main parts. In Section \emph{Survey Methodology}, we described the guidelines adopted in this work. Section \emph{Conceptual Scheme for Spatial Models} detailing the main spatial models found in the Bayesian spatial literature. Then, Section \emph{Analyses} shows the empirical results obtained from the meta-analysis, over the published papers in the last 20 years, and Section \emph{Concluding Remarks} enlightening the coming up step on the field.

\section*{Survey Methodology}\label{survey}

 The data collection focused on determining the field where Bayesian spatial statistics is most applied, the current development stage of spatial models, and to identify the contribution trend in Bayesian spatial literature. The collection and reporting methods were based on the guidelines of the Preferred Reporting Item for Systematic Review and Meta-Analysis (PRISMA) \citep{Hutton2015, Moher2009b}. This procedure includes an electronic search strategy, a clear objective to define the inclusion and exclusion criterion, and an appropriate method of reporting findings.

An online electronic search was conducted on June 10, 2020, in the following four databases: Elsevier's Scopus, Science Direct, Thompsom Reuters's Web of Science, and the American Mathematical Society's MathSciNet database. Queries of the word ``Bayesian Spatial'' and ``Bayesian spatial'', using the Boolean operator ``OR'', through the year 2001-2020 was conducted. Title, abstract, and Keywords, were used in Scopus and Science Direct, topic (which entails title, abstract, and keywords) in Web of Science, and ``Anywhere'' in MathSciNet. 

The time frame was chosen to capture the diversification of the application of Bayesian spatial models in various fields, Owing to the advancement in technology for data collection and computation. Mendeley Windows application was used to remove duplicated articles. The resulting set was further examined manually for more duplicates not identified by Mendeley's application. The titles and abstracts of articles included after removing duplicates were first screened for Bayesian spatial methodology before applying the following inclusion criteria.
\begin{itemize}
    \item Search results that are written in English, an article published in peer-reviewed journals available online  
    We exclude books, dissertations, thesis, conference proceedings, reviews (or any other form, not an article).
    \item Articles that specifically implement Bayesian spatial models excluding articles that only mentioned Bayesian spatial models.
\end{itemize}

\begin{center}
    \begin{figure}
        \centering
        \includegraphics[width=12cm,height=9cm]{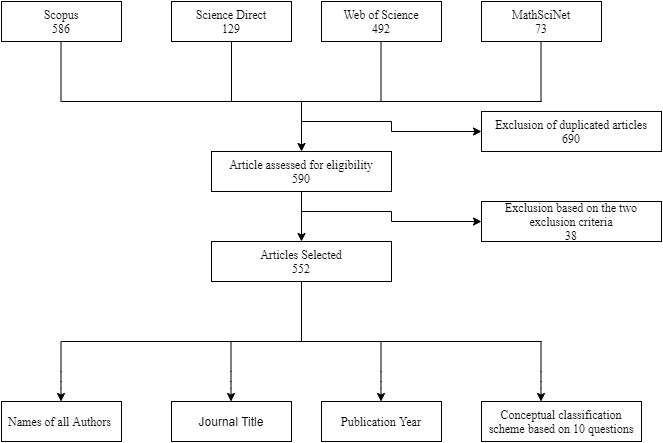}
        \caption{Flow chart of search procedure}
        \label{fig:flowchart}
    \end{figure}
\end{center}

Articles that did not meet the two inclusion criteria were excluded from the review. The search flow chart is presented in Figure \ref{fig:flowchart}. Using the search keywords earlier mentioned, 586 articles were retrieved from Scopus, 129 from Science direct, 492 from Web of Science, and 73 in MathSciNet. After the exclusion of duplicated articles, 590 articles were assessed for eligibility, and 38 were further excluded based on the two exclusion criteria, leaving 552 articles selected for conceptual classification.

As a structure of the data set, 552 articles that met the eligibility criteria were classified into the following categories. 
\begin{itemize}
    \item Names of all authors,
    \item Publication year,
    \item Journal title and
    \item Response to the ten items of the conceptual classification scheme on Bayesian spatial models.
\end{itemize}

This survey was divided into two parts \emph{theoretical models} and \emph{empirical analyses} of the published articles. These analyses scrutinized results from the last 20 years in the next section. In the first part, we discussed the different approaches under the statistical innovation and their differences, divided into \emph{eight topics}; Fields of application, Spatial domains, Spatial Priors, response variables, statistical models, Prior specification, estimation methods, Simulation, and validation.
We presented the various applications adopted by the existing spatial models from the systematic methodological review in part two.


\section*{Conceptual Scheme for Spatial Models}\label{scheme}
This research focused on content analysis in Bayesian Spatial model to systematically assess the content of a large volume of recorded information in this field. It aims to provide a deep insight into the contributions and identify the key research directions and opportunities in the Bayesian spatial methodology. To accomplish this, we applied a conventional approach to content analysis \citep{Hsieh2005} by scrutinizing through samples of the articles to clearly define the characteristics that better explain the scope and richness of the literature, identify the key concepts and patterns. This stands as the initial characteristics clustering. As we inspect the entire articles, subsequent updates on the clustering were included when new data that did not fit into the defined characteristics were encountered. This approach gives room for the literature to be classified without presumption.

Every Bayesian spatial analysis aims to estimate the spatial pattern over an extended geographical region to identify regions with extreme realization. In Bayesian hierarchical models, the spatial pattern is represented with a component that uses the same set of smoothing parameters across the entire study region. This type of smoothing is referred to as global smoothing. In some geographical setting, a global smoothing may be inappropriate, owing to the complexity of the geographical setting, and the spatial pattern is likely to exhibit a localized behavior. Thus, localized regions are smoothed using different parameters, and it is referred to as local smoothing \citep{lee2012}.


\begin{table}[htbp]
\caption{Summary of the Spatial Models and its variation.}
\resizebox{\textwidth}{!}{%
\begin{tabular}{|c|c|c|c|c|c|c|c|c|c}
\hline
 & & \multicolumn{ 2}{c|}{Spatial Smoothing} &\multicolumn{ 2}{c|}{ Gaussian Process }& \multicolumn{ 3}{c|}{Non-Gaussian Process } \\ \hline
\multicolumn{1}{|c|}{Spatial MODEL} & \multicolumn{1}{c|}{ARTICLE} & Global & \multicolumn{1}{c|}{Local} & \multicolumn{1}{c|}{GMRF } & \multicolumn{1}{c|}{Non-GMRF} & \multicolumn{1}{c|}{Prametric} &\multicolumn{1}{c|}{Semi-Parametric} & \multicolumn{1}{c|}{Non-Parametric} \\ \hline
CAR dissimilarity & Lee \& Mitchell, 2012  \cite{lee2012boundary} &  &\blackcheck&\blackcheck   & & \blackcheck & &\\ \hline
Intrinsic CAR/BYM & Besag et al., 1991 \cite{Besag1991}& \blackcheck &  &  \blackcheck & & \blackcheck  & &\\ \hline
Proper CAR  & Besag, 1974 \cite{Besag1974}& \blackcheck &  &  \blackcheck &  & \blackcheck  & &\\ \hline
Leroux & Leroux et al., 2000 \cite{leroux2000}&\blackcheck &  &  \blackcheck &  & \blackcheck  & &\\ \hline
Geostatistical & Clements et al., 2006 \cite{Clements2006}&\blackcheck&  &  \blackcheck &  &\blackcheck  & &\\ \hline
Globalspline & Lee and Durb{\'a}n (2009)\cite{lee2009smooth}  & \blackcheck &  &  & \blackcheck & & \blackcheck &\\ \hline 
Simpson CAR & Simpson $et~ al$ \citep{Simpson2017}   & \blackcheck&  &  \blackcheck  &&\blackcheck &  &\\ \hline
Dean's CAR &   Dean $et~ al$ \citep{Dean2001}  & \blackcheck&  &  \blackcheck  && \blackcheck&  &\\ \hline
SPDE & Lindgren, Rue \& Lindström, 2011 \cite{Lindgren2011} & \blackcheck&  &  & \blackcheck & \blackcheck & &\\ \hline
Mixture Model & Green and Richardson \citep{Green2002}  & \blackcheck&  &  & \blackcheck &  & \blackcheck&\\ \hline
Spatial Partition Model & Leonhard and Ra{\ss}er \citep{knorr2000bayesian}   & &\blackcheck  &  & \blackcheck &  &\blackcheck &\\ \hline
Asymmetric Laplace &  Kuzobowski and Pogorski \cite{Kozubowski2000} & \blackcheck&  &  & \blackcheck & \blackcheck & &\\ \hline
Student-t & Fonseca \citep{Fonseca2014}  & \blackcheck &  &  & \blackcheck & \blackcheck & &\\ \hline
Log-Gamma & Bradley $et\; al.$ \citep{Bradley2018}  & \blackcheck &  &  & \blackcheck & \blackcheck & &\\ \hline
Dirichlet  & Gelfand et al., 2005 \cite{gelfand2005bayesian}  & \blackcheck&  &  &\blackcheck &  & &\blackcheck \\ \hline

\end{tabular}}
\label{TAB:spatial_summary}
\end{table}
For instance, let's suppose as a special case of the spatial models, a general framework of the parametric spatial model given response variable $Y$, for instance, let's consider $\theta$ as a generalization of the response variable, and a set of a linearly related covariate is formulated as 
\begin{equation*}
    \bb{\theta}= g^{-1}(\textbf{X} \beta + \textbf{Z} \gamma),
\end{equation*}
where $\bb{\theta}$ is the transformed quantity of interest, $g(.)$ is a link function, \textbf{X} is the fixed effect design matrix, $\beta$ is the unknown fixed effects vector, \textbf{Z} is the random effect design matrix (particularly, the spatial effect), and $\gamma$ is the latent spatial variable to be estimated. The probability distribution of \textbf{Y}, $Y\sim f_{y}(.|\bb{\theta})$,  determines the function $g(.)$. In a frequentist estimation procedure, it involves maximizing the joint log-likelihood $l(y,\gamma|\beta)$. In a Bayesian framework,  all the parameters are considered random, which can be guided by an informative structure (prior). 

The spatial effect is assigned a spatial prior $\bb{\gamma}\sim f_{spat}(.|\bb{\phi},\Sigma(\alpha))$, where $\bb{\phi}$ is the vector of hyperparameters, and $\Sigma(\alpha)$ is a covariance matrix that determines the spatial dependencies of $\bb{\gamma}$ across a spatial domain with smoothing parameter $\alpha$. The $f_{spat}(.)$, usually multivariate,  assumes a parametric distribution (not necessarily an exponential family) such as Gaussian, Asymmetric Laplace, Student-t, Log-Gamma, and more. The $\alpha$ is a global smoothing parameter if the same set of $\alpha$ smooths the entire spatial region, whereas it is a local smoothing if different sets, represented by the vector $\bb{\alpha}$, smooth the spatial region. Additionally, the spatial process can be modeled in a semi-parametric framework, such as the spatial mixture model, and non-parametric, such as the Dirichlet process. 

In the review process, as a research methodology, we first classify each article into one of the two disjoint classes, "\textit{Theoretical}" and "\textit{Applied}". The theoretical methods involve investigating fundamental principles and reasons for the occurrence of an event, random phenomenon, or processes. On the contrary, applied research involves solving a particular problem with known or accepted theories and principles.

\subsection*{Spatial Statistics Fields of Application}%
Bayesian spatial statistics is a useful tool to determine the dependence structure and hidden patterns over space, through the prior knowledge and data likelihood. In some cases, the hypotheses of interest of a random phenomenon do not directly relate to the effect of spatial dependencies. However, it is crucial to adjust for spatial variation \citep{Riley2015}. Adjustment for spatial patterns in modeling random occurrence has since been practiced across various fields such as Agriculture, Medicine, Biology, Epidemiology, Geography, Geology, Economics, Climatology, Ecology, among others \citep{Karimi2012}. Moreover, spatial dependence in the Agriculture experiment has long received consideration. RA. Fisher identified spatial variations and used blocking to mitigate the effect of spatial dependencies in a randomized experimental design \citep{Mark2015, Fisher1935}. 

In many Biological and Medical experiments, such as gene classification, brain mapping, the randomized blocking technique may not be a viable alternative. Moreover, in demography, disease mapping, image analysis, remote sensing, fabrication engineering, and species detection, the variation due to spatial proximity cannot be neglected. It may result in bias and inconsistent estimates. Responses at close range tend to have similar behavior and variation. The homogeneity of the variation depreciates with the increased distance apart. An efficient procedure to tackle the effect of spatial proximity is to consider random field Statistical models. Random field Statistical models, known as spatial models, describe the distribution of a random phenomenon over a spatial domain.

Spatial models have long be applied in various fields. In 1949, Isard described the general theory of the spatial formation of economic activities focusing on the geographic distribution of costs, prices, and location of industries \citep{Isard1949}. Spatial statistics applied to economics, often referred to as spatial econometrics, have gained more attention in recent years to analyze economic data over a wide range of spatial domain \citep{Sparks2013}. Similarly, In 1950, DA. Krige took advantage of nearby variations to pursue the spatial prediction of gold distribution in South Africa, basing predictions practically on lognormal-de Wijsian spatial models \citep{Krige1978}. In Epidemiology and Public Health, spatial statistics have gained increasing importance in predicting disease outbreak \citep{Gracia2015, Luan2018, Morris2015, Mueller2002, Short2002}. The problems that arose from these fields lead to the motivation of several intuitions that gave birth to the consistent improvements in the spatial model literature. Some of the most advances are identifying spatial risk factors, disease surveillance, and spatial predictive models, \citep{Ward2008}.

In this research, the fields of the application were classified into five major parts. 1. Biological and Medicine: These include researches in Biology, Medicine, Epidemiology, and Public health, 2. Economics and Humanity: These include Economics, Demography, Criminology, Accident analysis. 3. Physical science and Engineering, 4. Agricultural and Environmental Science, and 5. Sport.

\subsection*{Spatial Domains}
Geographically reference data, also known as spatial data, is a collection of a stochastic process indexed by space. In other words, Suppose $Z(s)$ is a random process observed at location $s$, the set $Z(s) \equiv \{ z(s), s\in D\}$ is a spatial data, where $D$, a subset of $\field{R}^d$, usually ( but not necessarily ) fixed and represents a spatial domain. According to Blangiardo \citep{Blangiardo2015}, the spatial domain are distinguished as follows
\begin{itemize}
    \item Area or Lattice data: it is a simple way to represent spatial data in the domain $D$. In this type of spatial domain, $z(s)$ is a random aggregated realization across an area $s$ of distinct boundaries. For area data, the boundaries are irregular, such as administrative divisions, while for lattice, the boundaries are a regular division of $D$.  For simplicity purposes, it may be necessary to aggregate other types of Spatial domain realizations to form area or lattice data. This process may sometimes be referred to as a discretization of $D$.
    \item Geostatistical or Point-reference data:  $z(s)$ is a realization at a specific location $s$ in a continuous spatial domain $D$. The location $s$ is considered a coordinate made up of longitudes and latitudes, and sometimes includes altitudes. The location $s$ could also be represented in Cartesian coordinates.
    \item Spatial point pattern: The realization $z(s)$ represents the occurrence or non-occurrence of an event at location $s$. In this case, the location itself is considered to be random. The random realization is a location indicator of the presence or absence of a phenomenon of interest in the domain $D$. In Agriculture, for example, the interest may be the distribution of a specific tree species, where each realization is the presence or absence of the tree specie in domain $D$. In epidemiology, the realization may be the house address of a patient having a particular disease \citep{Banerjee2004, Cressie1993}.
    \end{itemize}

\subsection*{Spatial Priors}
It is necessary to determine or specify a prior distribution for the posterior distribution's complete estimation in an empirical or full Bayesian approach. For example, in hierarchical models, the prior distribution assumed for a random field ( spatial component ) $z(s)$ is termed, spatial model. We encountered several types of spatial models (priors) in the literature, and most were sub-class of the Gaussian Markov Random Field (GMRF) defined as a Gaussian random field with Markov property \citep{Cressie1993, Rue2002}. 

In the literature, due to the large class of priors, we collapsed the encountered priors based on the most frequently used and seldom-used classes, which are the \textit{Conditional Autoregressive (CAR), Beseg York Mollie (BYM), Lorex CAR, Stochastic Partial Differential Equation (SPDE), GMRF (none of the above), Non-GMRF and Others}. See table  (\ref{TAB:spatial_summary}) for summary. Details of each model are presented in the Appendix. The Lorex CAR class comprises priors with similar specifications, such as Dean's and Simpson CAR models. The GMRF class consists of GMRF priors except for the CAR family and the SPDE earlier stated. 

The non-GMRF is a large class that consists of non-trivial prior models that utilize a spatial correlation function to determine the covariance matrix of the spatial process in a continuous space, including the Asymmetric Laplace, Log-Gamma, skewed normal, Student-t process, and Dirichlet process.  We created a class $Others$ to accommodate unspecified models and those that do not belong to the above classes.


A response variable is a quantity used to describe a random process to relate it to a deterministic process mathematically. In statistical modeling, the most frequently used response variables are the discrete (categorical), ordinal, and continuous variables. The type of variables used in modeling a random phenomenon is intuitive from the process under study. The statistical models used to describe a random phenomenon vary depending on the quantity and parameters of interest. 

The Bernoulli distribution is often used for modeling the random phenomenon of two possible outcomes. The Binomial, Negative Binomial, Hypergeometric, and Poisson distribution are frequently used for modeling count cases such as disease occurrence, wildlife, signal, and more. The Poisson distribution has been used to approximate Binomial distribution for a large sample size \citep{teer2014,burr1973some}. The equality of mean to variance restriction imposed by the Poisson distribution considers the Negative Binomial a better choice to model a random variable that exhibits over-dispersion. The Multinomial distribution is often used to model a phenomenon of more than two categories usually encountered in Biological experiments. It is a generalization of the Binomial distribution. 

In the continuous case, a large class of distribution of the exponential family is used, such as Gaussian, Exponential, Student t, Weibull, Gamma, and more. However, according to the Central limit theorem, the Gaussian distribution is used to approximate both discrete and continuous distribution for large sample size \citep{kwak2017central}.

An analyst's interest is to quantify the association of a random phenomenon and explanatory processes, describing a random phenomenon according to a set of explanatory variables. In the literature, the statistical models encountered are the Generalized Linear Mixed Model and the Hierarchical model, \textit{Survival model}, and \textit{Spatial Econometrics models}. The details of each model are presented in the appendix.  Additionally, we created a \textit{Proposed} $Unspecified$ and $Others$ classes to accommodate proposed and validated models and unspecified models. The class of $Others$ accommodates statistical models outside the above listed classes.

An appropriate prior distribution specification in a Bayesian inference remains a challenge in various fields of application. A prior distribution is associated with the representation of uncertainty of the interest parameters before data are observed. The elicitation of an appropriate prior distribution is a non-trivial task \citep{palacios2006}, and such challenges are accumulated in spatial models due to the large number of associated parameters involved. In our review, we came across four main approaches.

One way to set a prior distribution is to assume ignorance about the appropriate model, that is, $\mathbb{P}(\theta\propto 1)$, and allows the data model to carry all the information. Such an approach is not always advantageous because inference on the parameters can be improved by performing prior elicitation based on identified characteristics or expert opinion. The elicitation procedure is termed elicited prior.

In elicited priors, convenient prior distributions are sometimes a choice and have spread across literature and have been set as default priors in most simulation packages. As a result, subsequent authors use such prior distribution verbatim. However, several authors did not explicitly state the type of prior used and were classified as not available.

\subsection*{Estimation Method}
In Bayesian inference, the prior information expressed through the prior distribution $\mathbb{P}(\theta) $ and the data likelihood $\mathbb{P}(y|\theta) $ are used for inferences. In simple or convenient cases, the posterior distribution given in (\ref{Posterior}) is easy to evaluate. However, in practice, evaluation of the posterior distribution to make inferences is a non-trivial task because it usually contains compound integrands with complicated support that is not analytically integrable \citep{Fragoso2018}. In this sense, authors explore different approaches to make inferences. In the literature, we encountered several estimation methods and classify them into the \textit{Markov Chain Monte Carlo (MCMC), Integrated Nested Laplace Approximation (INLA), Expectation-Maximization (EM) and Maximum (Penalized quasi) Likelihood Method classes}. Details of these estimation methods are presented in the appendix. The MCMC class comprises of all numerical approximation that utilizes Monte Carlo method. Also, the \textit{unspecified} class was added to accommodate articles that neither discuss nor state the approach used in the estimation procedure. The \textit{Others} class comprises of estimation methods that do not fit into the defined classes. 
\begin{align}
\begin{aligned}
     \mathbb{P}(\theta|y)=\frac{\mathbb{P}(y|\theta)\mathbb{P}(\theta)}{\int_{A} \mathbb{P}(y|\theta)\mathbb{P}(\theta) dy}\\
     \mathbb{P}(\theta|y) \propto \mathbb{P}(y|\theta)\mathbb{P}(\theta).
    \label{Posterior}
\end{aligned}
\end{align}

\subsection*{Simulation Study and Validation}
A simulation study is a systematic and scientific computer procedure that involves fixing model parameters to generate data by pseudo-random sampling \citep{morris2019}. It comprises two main steps; the data generation and the estimation. In the first step, a set of parameters is fixed and used to generate pseudo-random data. In the second step, the generated data is fed back to the model to estimate the "unknown" parameters and check for bias and model error. 

A simulation study is usually carried out for proposed models and methods. The articles reviewed were classified into two; "Yes," if the article contains statistical simulation studies, and "No" if it does not.
 
In addition to the simulation studies, we also investigated how Bayesian spatial models were validated using real data. It is a procedure to test for overfitting or underfitting. A model overfits if it performs well in the training set and badly in the test set, while it underfits if it performs poorly in the training set. A classical approach to cross-validation is to form a disjoint subset of a whole data into training and testing sets. The model is fitted on the former and tested on the later set. Doing this process $ k $ times until all observations in the data set participate in training and testing, once, it is called K-fold cross-validation. The whole data is split into $k$ disjoint subsets, where the combined $k-1$ sets serve as the training set and the remaining set of size $n_k=n/k$ serve as the test set $n$ is the data size. A particular case to the k-fold is the Leave-One-Out cross-validation, where one observation serves as the test set, and the remainder $n-1$ serves as the training set. After going through all subsets, the validation measures are statistically combined to make a valid conclusion.

Since the spatial models are often modeled in a Bayesian framework, we included the \textit{Posterior predictive check} \citep{gelman1996} class. In a predictive posterior check, a test statistic is chosen and computed for the observed data process. The same statistic is computed for replicated posterior predictions of the process. The model is said to present a good fit if the posterior prediction average is close to the test statistic for the observed data \citep{Fragoso2018}. We added \textit{None or not applicable} class to accommodate articles that did not conduct cross-validation and \textit{Others} to accommodate the validation method not mentioned above.
 

\section*{Analyses}\label{empirical}

As described in the search procedure section, a total of 552 articles were selected after filtering with the exclusion criteria (duplication and context). After a careful watch, the articles were categorized into the first class, labeled as being only an applied paper, theoretical, or both, where 4 (1\%) of the papers showed no application (only theoretical with synthetic data), 188(34\%) showed an improvement in the field with real-world application and 360(65\%) only applied the existing methodologies.

The papers were sub-divided into five classes of application field: \emph{Agricultural and Environmental Science}, \emph{Economics and Humanities}, \emph{Medical Science}, \emph{Physical Science and Engineering} and \emph{Other}. Mainly, three fields hold the majority of the publications, which are \emph{Agricultural and Environmental Science} (30.1\%), \emph{Economics and Humanities} (30.6\%), and \emph{Medical Science}, which includes epidemiology, (33.7\%). Moreover, the spatial domain used was also taken into account. The \emph{Area/lattice}, \emph{Geostatistical}, and \emph{Spatial Point Patterns}. 
The \emph{Area/lattice} occurred 65.6\% and \emph{Geostatistical} 31.2\% of the reviewed articles, and in combination, they hold 95.8\% of the publications. It is important to note that more than one spatial domain could be used in an article, such as the 1\% observed in this review. This procedure is common when a continuous spatial domain is discretized to lower computational burden. 

The core part of this review is the spatial priors (models) used in the literature. While the literature contains numerous spatial priors applied in various problems across various research fields, this systematic review only presented models encountered during the review. Gelfand et al. \cite{gelfand2010handbook} explicit the first incorporation of a spatial prior, and it is gain.

Table \ref{TAB:spatial_summary} summarized the spatial models encountered and were classified into distinguishable groups. Among the spatial models, the \emph{CAR family}, usually used for area/lattice spatial domain, \emph{Stochastic Partial Differential Equations (SPDE)}, \emph{Gaussian Markov Random Field (GMRF)} except for the CAR family, models with author-specific defined covariance structure often for a continuous spatial domain, and Non-Parametric methods. The CAR family appeared in 44.2\% of the published articles reviewed. Of this percentage, the CAR and the BYM model appeared 96.3\%. The SPDE appeared about 3.9\%, and the non-parametric procedure appeared 1.3\%  of the articles reviewed. The GMRF, exempting the CAR family and the SPDE, appeared in the literature 4.0\%. Consequently, to the diverse application of the spatial model, authors specified the type of covariance structure based on the prior knowledge of the interactions between the phenomenon of interest that appeared 31.3\% of the articles reviewed. Only a single documentation of the application of spatial models on robotic technology was encountered \cite{taniguchi2017online}. Other models that could not fit into any of these groups appeared 7.2\%, and 9.3\% of the articles did not described or state the type of model adopted. 

\begin{figure}
    \centering
    \includegraphics[scale=0.5]{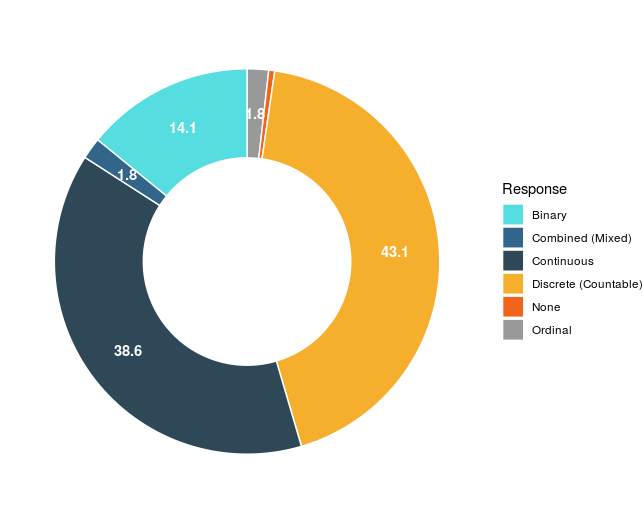}
    \caption{Distribution of the response variable type.}
    \label{fig:res_type}
\end{figure}
\begin{figure}
    \centering
    \includegraphics[scale=0.4]{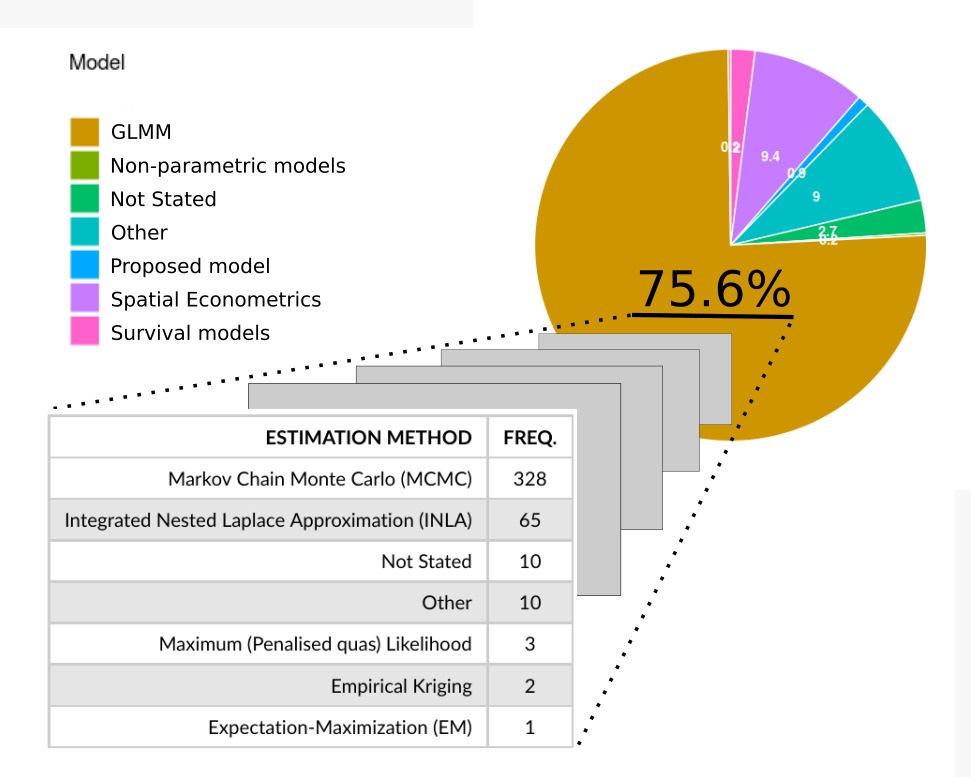}
    \caption{Class of models often used the spatial modelling and its numerical estimation method distribution. Given the class of GLMM/Hierarchical Models, Markov Chain Monte Carlo (MCMC) is the most used computation intensive technique.}
    \label{fig:inf_model}
\end{figure}

The observation or response variable's nature dictates the statistical model class to be adopted to make inferences. In our search, as shown in Figure \ref{fig:res_type}, the discrete (Countable) was the most used, then the continuous and binary response variable types. 
Additionally, the most adopted statistical model to spatial analysis, withing a Bayesian framework, is the GLMM. Owing to the integral complexity of the posterior marginal distribution, the MCMC estimation method is the most frequently adopted numerical integration, Figure (\ref{fig:inf_model}). 

Maybe the most critical question regarding Spatial Models is related to specifying Spatial Prior. Elucidating events in an area, some times even under a few frequencies, is desirable through direct probabilistic statements that may unravel hidden patterns \cite{kang2016making}. Whereas the inter-dependence may conduct spatial correlation with Bayesian Models across the spatial fields, and the Bayesian approach enables the information expertise to be allocated with the acquired data. The results obtained in this systematic review, as Table \ref{tab:prior} shows that the knowledge of the expert is used (30.43\%), although it can be better explored.
\begin{table}[htbp]
\centering
\caption{Model Prior specified}
\begin{tabular}{r|c}
\hline
\textbf{Prior specified} & \textbf{Freq.} \\ \hline
Elicitated from experts or from the problem  & 168 \\ 
No explicit use or reference/not applicable  & 101 \\ 
          Used verbatim from the literature  & 166 \\ 
              Vague prior (Non-informative)  & 119 \\ \hline
\end{tabular}
\label{tab:prior}
\end{table}

\begin{figure}
    \centering
    \includegraphics[scale=0.4]{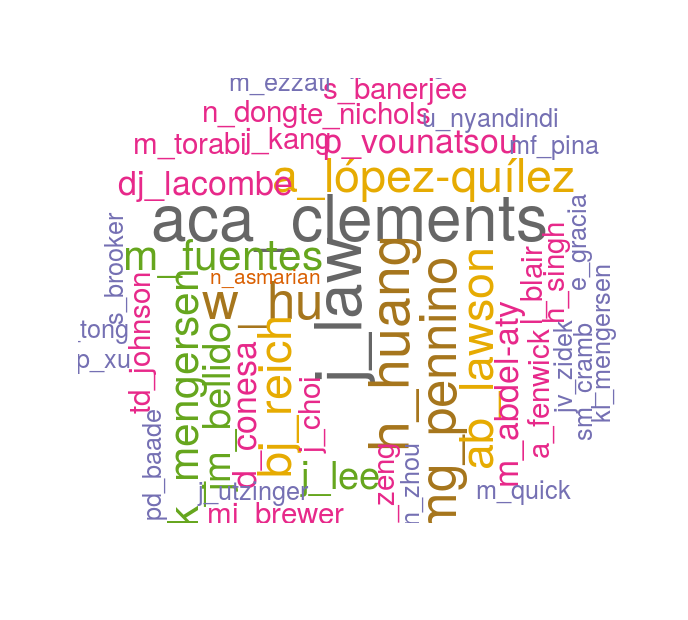} \quad
    \includegraphics[scale=0.35]{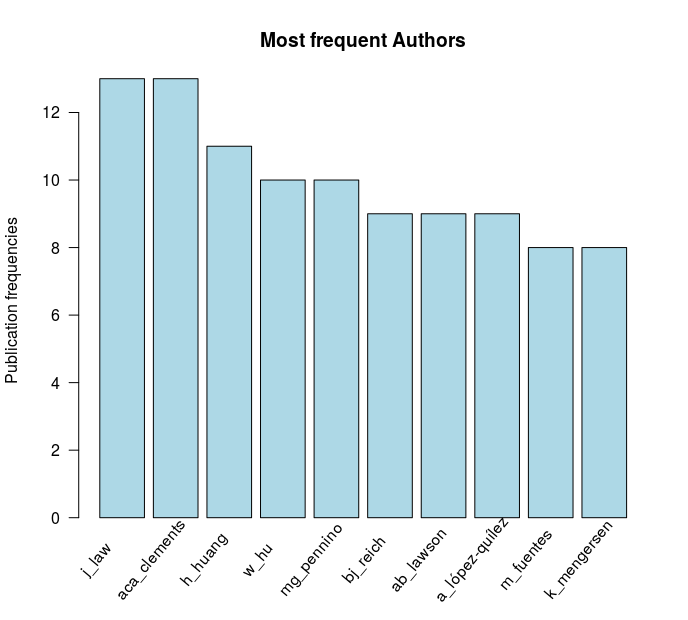}
    \caption{Most frequent authors on the Bayesian Spatial Models. Left-hand graph is a Tag Cloud for the 50 most frequent authors in the past 20 years. Right-hand graph is a Barplot displaying the Top 10 authors and its relative frequencies.}
    \label{fig:authors}
\end{figure}

The Authors who were more recurrent in the literature in this past 20 years were James Law, ACA Clements, and Hei Huang. Figures \ref{fig:authors} displays the visualization towards the relevant authors in the Bayesian Spatial Models publication field.

\begin{figure}
    \centering
    \includegraphics[scale=0.4]{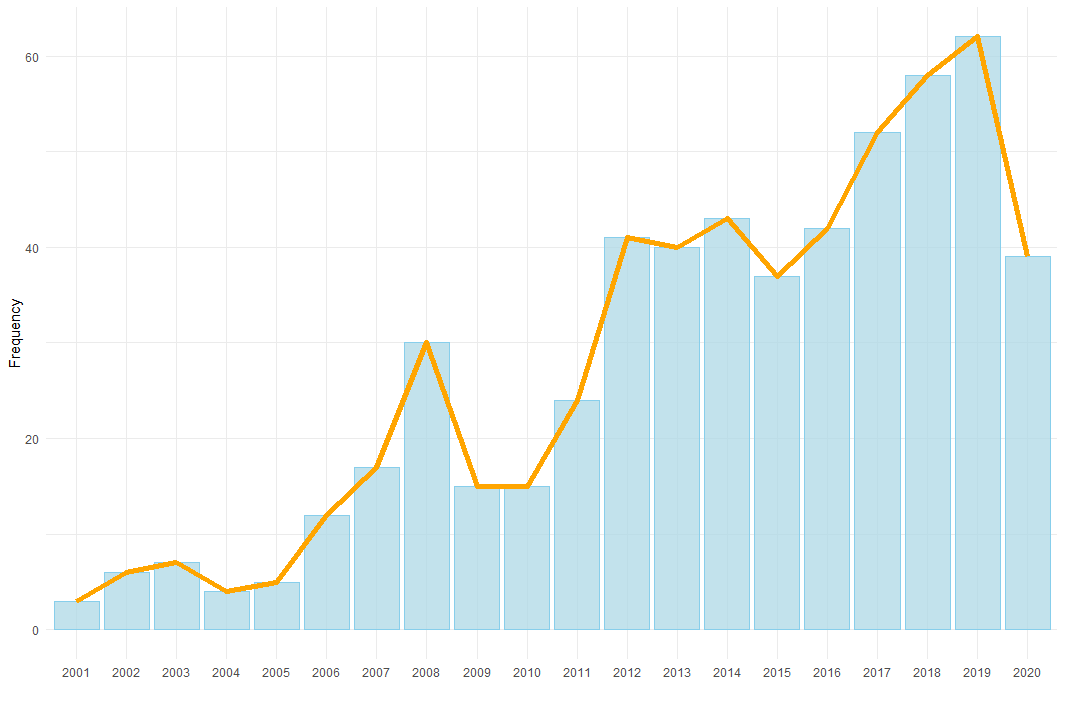}
    \caption{Publication Growth, regarding the Bayesian Spatial Models, across time.}
    \label{fig:time}
\end{figure}

The top 5 journals which most contained articles, from the obtained data set, which combined theoretical methodology with real-world application publishing on Spatial and Spatio-temporal Epidemiology (\#15), Accident Analysis and Prevention (\#14), PLoS ONE (\#14), Spatial Statistics (\#11), and Environmentrics (\#10). Whereas, across time, the spatial modeling publication rate using the Bayesian approach proliferates, as shown in Figure \ref{fig:time}. The year 2020 refers only to the first half of the year. 

\section*{Concluding Remarks}\label{conc}

Spatial statistics has gained tremendous attention in recent years owing to the efficiency in collecting spatial dependence data. Neglecting such dependencies may result in bias and consequently leads to wrong inference. The Bayesian approach to analyzing spatial data often outperforms the frequentist approach, given that the prior information is taking into account. In a Bayesian framework, spatial priors play a significant roll in accounting for space dependencies. The consistency in the improvement of data collection and computational tools in analyzing spatial data, Bayesian spatial statistics will further penetrate numerous fields and becomes one of the leading tools for analyzing data. 

Many authors account for spatial dependencies assuming a Gaussian random field. In many real data applications, the Gaussian random field may be inappropriate, especially in extreme data, skewed data, data with spikes and heavy tails. Examining a different random field such as the Laplace, Student-t, Pareto, Nakagami-m, and more may improve inference. A significant complication of assuming these distributions is the non-trivial method of fixing a prior distribution for the model parameters. For instance, the prior distribution for the degrees of freedom of the student-t distribution is not a trivial task \cite{ordoez2020objective}. Regardless of the prior distribution, eliciting priors for the parameters is critical, and when wrongly assumed, it could lead to misleading results and inference. To circumvent these, which is also not a trivial task, it is essential to consider objective priors for the random field parameters and hyper-parameters to improve inference. 

The Bayesian spatial literature lacks sufficient information on the objective priors, such as Jeffery's prior, reference priors, matching priors, and more. These priors stand out to elicit experts ideas that could improve the inferences of the spatial dependencies. To derive an objective prior distribution for a spatial random field's smoothing parameter is currently an open problem that needs urgent attention. 

The Markov property is a useful tool to lower the computational cost in Bayesian inferential statistics by subjecting the immediate neighbors' spatial dependencies. However, the realization of some random phenomenon exhibits strong spatial dependencies beyond the immediate neighbors, and truncating such dependencies structure will result in bias and incorrect inferences. Moreover, there is an insufficient standard approach to determine the covariance matrix structure of the spatial effects. Thus, spatial smoothing is not trivial to compare across different models.

In Neuroscience, the application of Bayesian spatial statistics to brain experiments is gaining interest \cite{derado2013predicting,kang2011meta,song2019potts,taschler2014spatial}. The complexity of the brain structure has prevented the application of classical spatial models. In other words, the primary assumption of spatial contiguity in the analysis may result in incorrect inference owing to brain complexity. Moreover, a response received at one location on the skull may be due to brain activity in the opposite location. Beyond complexity, the dynamics of the body system of the subject influences the experiments. Thus, accounting for such complex structures is an open problem that requires further studies. 

Despite the increasing availability of the Big IoT data, the spatial model has not received adequate attention as a classification algorithm and regression machine learning models. Application of machine learning to spatial data could unravel hidden patterns, and applied to efficiently navigates a robotic technology \cite{taniguchi2017online} through space, which creates new research lines.

\section*{Conflict of interest}

{\label{644442}}
There is no potential conflict involved in this report

\selectlanguage{english}
\bibliography{converted}

\section*{Appendix A}
The coding of the characteristics was framed according to the conceptual classification scheme developed by Hachicha and Ghorbel \cite{Hachicha2012} and applied by Tiago $et al.$ \cite{Fragoso2018}. In this framework, bias is limited in survey data and clarifies reporting results and findings to draw concrete conclusions. Such classification is useful to researchers as it provides an overview of the research methodology's application and can reveal research gaps in the literature.

This review employed a conceptual classification scheme of 10 items presented in Table \ref{tab:my_label}. For lucidity, each item in the classification scheme is discussed in detail.

\begin{longtable}{ll}
    \centering
      &List of questions for conceptual classification scheme \\
     \hline
    &1.0 Is it only an application?\\
    &\hspace{0.5cm}1.1 Yes\\
    &\hspace{0.5cm}1.2 Both\\
    &\hspace{0.5cm}1.3 No (only method)\\
      &2.0 What is the field of application? \\
     &\hspace{0.5cm}2.1  Medical science\\
     &\hspace{0.5cm}2.2 Economics and Humanity\\
     &\hspace{0.5cm}2.3 Physical Science and Engineering\\
     &\hspace{0.5cm}2.4 Agricultural and Environmental Science\\
     &\hspace{0.5cm}2.5 Sport\\
     &3.0 What Spatial domain was employed?\\
      &\hspace{0.5cm}3.1 Area or lattice\\
      &\hspace{0.5cm}3.2 Geostatistical data\\
      &\hspace{0.5cm}3.3 Spatial point patterns\\
      &\hspace{0.5cm}3.4 Area and Geostatistical data\\
      &4.0 What type of spatial priors used?\\
      
      &\hspace{0.5cm}4.1 Conditional Autoregressive (CAR)\\
      &\hspace{0.5cm}4.2 Besag York Mollié (BYM)\\
      &\hspace{0.5cm}4.3 Leroux CAR\\
      &\hspace{0.5cm}4.4 Gaussian Markov Random Field (Other specifications)\\
      &\hspace{0.5cm}4.5 Covariance Function (Not GMRF)\\
      &\hspace{0.5cm}4.6 Other (new methodology/proposed)\\
    &5.0 What type of response variable?\\
    &\hspace{0.5cm}5.1 Discrete (Countable)\\
    &\hspace{0.5cm}5.2 Continuous\\
    &\hspace{0.5cm}5.3 Combined (Mixed)\\
    &\hspace{0.5cm}5.4 Ordinal\\
    &6.0 What is the statistical model used?\\
     &\hspace{0.5cm}6.1 Generalize Linear (mixed) model (or Hierarchical models)\\
     &\hspace{0.5cm}6.2 Survival and Longitudinal models\\
     &\hspace{0.5cm}6.3 Non-parametric models (Machine Learning models)\\
     &\hspace{0.5cm}6.4 Spatial Econometrics\\
     &\hspace{0.5cm}6.5 Proposed\\
     &\hspace{0.5cm}6.6 Not stated\\
     &\hspace{0.5cm}6.7 Other\\

    &7.0 How are model Prior specified?\\
    &\hspace{0.5cm} 7.1 Vague prior (Non-informative)\\
    &\hspace{0.5cm}7.2 Used verbatim from the literature\\
    &\hspace{0.5cm}7.3 Elicitated from experts or from the problem\\
    &\hspace{0.5cm} 7.4 No explicit use or reference/not applicable\\
    &8.0 What is the estimation method applied?\\
    &\hspace{0.5cm}8.1 Markov Chain Monte Carlo (MCMC)\\
    &\hspace{0.5cm}8.2 Integrated Nested Laplace Approximation (INLA)\\
    &\hspace{0.5cm}8.3 Expectation-Maximization (EM)\\
    &\hspace{0.5cm}8.4 Maximum (Penalized quasi) Likelihood Method\\
    &\hspace{0.5cm}8.5 Not stated\\
    &\hspace{0.5cm}8.6 Other\\
    &9.0 Is the model validated through simulation?\\
    &\hspace{0.5cm}9.1 Yes\\
    &\hspace{0.5cm}9.2 No\\
    & 10.0 Is the application validated through data-driven procedures?\\
    &\hspace{0.5cm}10.1 Cross-validation and data splitting (K-fold / Holdout)\\
    &\hspace{0.5cm}10.2 Leave-One-Out Cross-Validation (LOOCV)\\
    &\hspace{0.5cm}10.3 Posterior predictive check\\
    &\hspace{0.5cm}10.4 Other\\
    &\hspace{0.5cm}10.5 None or not applicable\\
    \hline
    \caption{Classification scheme}
    \label{tab:my_label}
\end{longtable}

\section*{Appendix B}
In this Appendix, we made an overview of the primary spatial statistical models explored in this systematic review. 

\subsection*{Class of Gaussian Markov Random Field (GMRF)}
In standard form, the covariance matrix of a Gaussian random field is positive definite and often dense. This restriction makes it difficult to construct an appropriate covariance matrix. Furthermore, the computational cost is high due to the cost of $O(n^3)$ to factorize the covariance matrix \citep{Lindgren2011}. To overcome the above challenges, it is necessary to construct a sparse covariance matrix of less computational cost and closely approximate the dense covariance matrix. A possible choice is to adopt a Markovian property on the covariance matrix to have a GMRF approximation.
It is a discreetly indexed finite random vector that has a joint Gaussian distribution with a sparse precision matrix. Following the definition given by Rue \citep{Rue2005}, the set $Z(s) \in \field{R}^d$ over a spatial domain $s \in D$ is said to be a GMRF with respect to graph $(V, D)$, centered at $\mathbf{\mu}$ and precision matrix $\mathbf{Q>0}$ and density of the form
\begin{align}
    \pi( \mathbf{Z}| \mathbf{\mu,Q})=(2\pi)^{-n/2}|\mathbf{Q}|^{1/2}\exp \Big (\frac{-1}{2}(z-\mu)^{T}\mathbf{Q}(\mathbf{z-\mu})\Big ),
\end{align}
where $Q_{i,j}\neq 0$ if $i$ and $j$ are considered neighbors, $V$ is a set of vertices, and $\xi$ is the edges of graph G representing $D$. Usually, the covariance function is constructed such that it is a function of the relative positions of location $i$ and $j$ in $D$. $\mathbf{Q}$ is considered stationary when it is only a function of the distance between the locations in $D$.  Let $\delta_i=\{z:\; z \; \text{is a neighbor of}\; i\}$, the full conditionals is given as
\begin{align}
    \pi (z_i|\mathbf{Z}_{-i})=\pi(z_i|\delta_i),
\end{align}
for every $z_i \in \mathbf{Z}$, and $\mathbf{Z}_{-i}$ is the vector $\mathbf{Z}$ excluding $z_i$.
Some of the subclasses of GMRF identified in the literature for spatial smoothing are the proper and Intrinsic Conditional Autoregressive (ICAR), Beseg York Mollie (BYM and BYM2), Lourex CAR (Leroux CAR), and Dean's CAR.

\textbf{--Conditional Autoregressive (CAR)}\\
Let the spatial domain $D$ be partitioned into $n$ disjointed areas $E$ such that $D=\cup_{i=1}^{n} E_i$. $E$ could be regular or not, as in Lattice or area respectively.  Let $z_i$ be a random variable observed at $E_i$ and the vector $\mathbf{Z=(z_i,...,z_n)^T}$ with $\mathbf{\mu}=(\mu_1,...,\mu_n)$. Let $\mathbf{Z_{-i}}$ be an $(n-1)$ dimensional vector such that $\mathbf{Z}_{-i}=(z_1,...,z_{i-1},z_{i+1},...,z_n)^T$, the full conditionals of a proper CAR model is given as
\begin{align}
    \pi(z_i|\delta_i,\sigma^2,\mu_i) \sim N \Big (\mu_i+\sum_{\{j:z_j\in \delta_i\}}c_{ij}(z_j-\mu_j),\kappa_i^{-2} \Big ),
    \label{CAR_model}
\end{align}
where $\delta_i=\{z:\; z\in E_j \; \text{is a neighbor of}\; E_i\}$,$\;\kappa_i^{-2}>0\; \forall~i=1,...,n$ and $c_{ij}$ is a function of the adjacency between $E_i$ and $E_j$. To make $\pi(\circ)$ a proper distribution, $\mathbf{C}=(c_{ij})$ and $\kappa$ are carefully chosen. A frequent choice for $\mathbf{C}$ is a function of contiguity between the areas. Let $\mathbf{A}=(a_{ij})$ be an $n\times n$ matrix, such that  $a_{ii}=0$, $a_{ij}=1$ if $z_j \in \delta_i$ for $i\neq j$, and $0$ otherwise. Noticed that $\mathbf{A}$ is symmetric, since if $E_j$ is a neighbor of $E_i$, then $E_i$ is a neighbor of $E_j$. We define $c_{ij}=\rho \frac{a_{ij}}{d_i}$,  $\kappa^2=\frac{d_i}{\sigma^2}$ and $d_i=\sum_{j}a_{ij}$ \citep{Besag1974,Freni2018,Rue2005}. The full conditionals and the joint distribution of (\ref{CAR_model}) is given by
\begin{align}
\begin{aligned}
    &\pi(z_i|\delta_i,\sigma^2,\mu_i) \sim N \Big (\mu_i+\rho \sum_{\{j:z_j\in \delta_i\}}\frac{a_{ij}(z_j-\mu_j)}{d_i},\frac{\sigma^2}{d_i} \Big )\\
    &\mathbf{Z}\sim N_n \Big(\mathbf{\mu},\sigma^2(\field{I}-\mathbf{C})^{-1}\field{M} \Big ),
    \end{aligned}
    \label{CAR_model2} 
\end{align}
where $\field{M}=diag(\frac{1}{d_1},...,\frac{1}{d_1}).$ and $|\rho|<1$ is chosen such that $(\field{I}-\mathbf{C})^{-1}>0$. Beseg, York and Mollie \citep{Besag1991}, proposed the intrinsic CAR (ICAR), by setting $\rho=1$, then ${c_{ij}}=\frac{a_{ij}}{d_i}$. The full conditionals of the ICAR
\begin{align}
    \pi(z_i|\delta_i,\sigma^2,\mu_i) \sim N \Big (\mu_i+\frac{\sum_{\{j:z_j\in \delta_i\}}a_{ij}(z_j-\mu_j)}{d_i},\frac{\sigma^{2}}{d_i} \Big ),
    \label{ICAR_model}
\end{align}
The model inferred that the conditional expectation of $z_i$ equals the weighted deviations of its neighbors in addition to its mean.

\textbf{--Beseg York Mollie (BYM)}\\
The BYM, proposed by Beseg \citep{Besag1991}, is a variant of the CAR model that incorporates an additional term to control for overdispersion in spatial data. Suppose $z$ is partitioned such that $z_i=u_i+v_i$, the unstructured term $v$ is modeled as
\begin{align}
    v_i \sim N(\mathbf{0}, \sigma^2_v),
    \label{unstru}
\end{align}
and the structured component is modeled as $u_i \sim ICAR.$
Thus,
\begin{align}
    Var(z|\sigma_u,\sigma_v)=\sigma^2_v\field{I}+\sigma^2_u(\field{I}-\mathbf{C})^{-1}\field{M}.
\end{align}

The BYM poses identifiable problem such that each observation is represented by $u_i$ and $v_i$, thus, only the sum $u_i+v_i$ is identifiable \citep{Lee2011}. Setting appropriate hyperparameters is challenging. However, constraining $\phi$ to sum to zero allows the confounding problem to be avoided and both components to be fitted.

\textbf{--Dean's Conditional Autoregressive}\\
A reparameterized version of the BYM proposed by Dean $et~ al$ \citep{Dean2001} and its covariance matrix are given as
\begin{align}
\begin{aligned}
    \mathbf{z}=\sigma(\sqrt{\phi }\mathbf{u}+\sqrt{1-\phi}\mathbf{ v}),\\
    Var(\boldsymbol{z}|\sigma_u,\sigma_v)=\sigma^2((1-\phi)\field{I}+\phi(\field{I}-\mathbf{C})^{-1}\field{M}),
\end{aligned}
     \label{Dean_model1}
\end{align}
where $\sigma_u=\sigma^2 \phi$ and $\sigma_v=\sigma^2 (1-\phi)$.

\textbf{--Simpson Conditional Autoregressive}\\
For the purpose of scaling and interpretablility of the hyper prior, Simpson $et~ al$ \citep{Simpson2017}  proposed a modification of the BYM which avoid the problems possed by BYM model. The combined random effect  and the covariance matrix are given by,
\begin{align}
\begin{aligned}
      \mathbf{z}=\sigma(\sqrt{\phi}\mathbf{u_*}+\sqrt{1-\phi}\mathbf{ v}),\\
    Var(\boldsymbol{z}|\sigma_u,\sigma_v)=\sigma^2((1-\phi)\field{I}+\phi \mathbf{Q}),
    \end{aligned}
     \label{BYM2_model1}
\end{align}
where $\sigma \geqslant 0$, $\phi \in [0,1]$ and $\mathbf{Q}$ is the ICAR covariance matrix such that $Var(u_i) \approx 1$.

\textbf{--Leroux Conditional Autoregressive (Lourex CAR)}\\
Leroux $et al.$ \cite{leroux2000} proposed a variant of the CAR model which, unlike the BYM, only requires a single set of spatial component. The full conditionals is given by
\begin{align}
    &\pi(z_i|\delta_i,\sigma^2,\mu_i) \sim N \Big (\mu_i+ \frac{\rho\sum_{\{j:z_j\in \delta_i\}}a_{ij}(z_j-\mu_j)+(1-\rho)\mu_0}{\rho d_i+1-\rho} ,\frac{\sigma^2}{\rho d_i+1-\rho} \Big ),
    \label{lourex_CAR}
\end{align}
which is a limiting case of the $ICAR$ model. The model avoids the identifiability problem in the BYM model by the specification of a single hyper parameter for random effect \citep{Cramb2017}. The specification (\ref{lourex_CAR}) smooths the neighboring random effect weighted by $\rho$ and the global mean weighted by $1-\rho$.\\
\textbf{--Conditional Autoregressive dissimilarity}\\
 The variations of the CAR model earlier discussed exhibits a global degree of spatial smoothing. In many instances, a global smoothing may be inappropriate, especially areas that exhibit locally constrained spatial structure.  Lee and Mitchell \cite{lee2012boundary} proposed a  CAR dissimilarity to smooth area elevated risks which, depends on local spatial parameters. 
It is one among many proposed local spatial smoothings, such as locally adaptive model \cite{lee2013locally}, localized conditional autoregressive \cite{lee2014bayesian}, to list a few.  \\
\subsection*{Class of Gaussian Non-Markov Random Field Models}
Markov property is known to lighten the burden of factorizing a dense covariance matrix.  However, a random realization with a strong correlation structure outside its neighborhoods can be poorly accounted for. Also, the discretization imposed by the GMRF sometimes does not account for the presence of discontinuities in spatial domains. These could be addressed by relaxing the Markov assumption and possibly allowing a different distribution other than Gaussian.\\
\textbf{--Spatial Weight Matrix}\\
In spatial Econometrics, a weight matrix and a correlation parameter are often used to specify a dependency structure between economic variables of interest observed at different spatial locations in a Spatial lag model (SLM) and Spatial Error Model (SER) \citep{griffith2018gis}. The weight matrix is essential in the covariance matrix specification for parametric models, such as the class of GMRF prior. The spatial weight configuration specification varies with the relationships between geographical locations, such as spatial distance, interactions, nearest neighbors, and contiguity. The specification is categorized into four: adjacency-based weights, weight-based on geographical distance, the distance between covariate values, and hybrid of geographical distance and covariates \citep{duncan2017spatial}. In most cases, nearby neighbors to a spatial reference location receive higher weights compared to farther neighbors. Though the diagonal elements of the spatial weight matrix are universally accepted to be set to zero \citep{griffith2012selected}, in the literature, there is no standard to define the weight matrix the specification is dominated by choice of computational convenience \citep{griffith2018gis}. \\
\textbf{--Spatial Covariance Function}

In this section, we briefly discussed the frequently used stationary and isotopic correlation functions to construct a valid correlation matrix for a random field in a Geostatistical models. A covariance function $C(s_i,s_j)$ is said to be stationary when $C(s_i,s_j)=C(s_i+l,s_j+l)$  for any lag $l$, and isotopic when it only depends on the distance $s_i$ and $s_j$.  A covariance function must be positive definite and symmetric.\\
Consider a Gaussian random field $Z(s)$, such that the realizations $\{z(s_i),s_i \in D,i=1,...,n\},$ $D\subseteq \field{R}^2$ are of interest. Note that the information we seek can be described by the mean $\field{E}(Z(s))$ and the covariance matrix $Cov(s_i,s_j)$. Thus, we could describe the mean by a linear function of covariates, and the covariance matrix as $Cov(s_i,s_j)=\sigma^2 \rho(||s_i-|s_j|),$ where $\rho(||s_i-|s_j|)$ is the correlation function.\\
Let $d(s_i,s_j)$ be Euclidean distance from site $i$ to $j$. The four most commonly used correlation function \citep{Cressie1993,Rue2002} are given by,
\begin{align}
\begin{aligned}
    Exponential:~ \rho(s_i,s_j)&=exp \Big (-3\frac{d(s_i,s_j)}{r}\Big ),\\
    Gaussian:~ \rho(s_i,s_j)&=exp \Big (-3\frac{d(s_i,s_j)^2}{r^2}\Big ),\\
    Spherical:~\rho(s_i,s_j)&=
    \begin{cases}
    1-\frac{2}{\pi} \Bigg (\frac{d(s_i,s_j)}{r}\sqrt{1-\Big (\frac{d(s_i,s_j)}{r} \Big )^2}+sin^{-1}(\frac{d(s_i,s_j)}{r} \Bigg ),& ~~ d(s_i,s_j) \leqslant r,\\
    0,& d(s_i,s_j) > r
    \end{cases}\\
    Matern: ~~\rho(s_i,s_j)&=\frac{1}{\Gamma(v)2^{v-1}}\Bigg ( S_v \frac{d(s_i,s_j)}{r} \Bigg )^v K_v \Bigg ( S_v \frac{d(s_i,s_j)}{r} \Bigg ),
    \end{aligned}
    \label{Corr_function}
\end{align}
where $K_v$ is a modified Bassel function of the second kind of order $v$, $S_v$ is the scaling factor, and $r>0$ defines a significant range of correlation for exponential, Gaussian, and Matern correlation function. While in spherical correlation function, $r$ is defined as the correlation length  \citep{Rue2002}, and $v$ is the smoothness parameter, which measures the differentiability of the Gaussian random field \citep{Sun2008}.

\textbf{--Skewed Gaussian random field}\\
Skewed form of the Gaussian random field has been used to model a skewed air pollution data. A skewed Gaussian random field cannot be defined in the same way as a Gaussian random field due to its marginals' dependence on its component parameter. For condense information, see \cite{Karimi2011,Karimi2012,Rivaz2016}.

\textbf{--Geostatistical model}\\
The geostatistical model proposed by Clement $et~al.$ incorporates an exponential distance decay on the average of a Gaussian process in a geostatistical model given by
\begin{align}
\begin{aligned}
    z_i&\sim N(\mu_i,\sigma^2),\\
    \mu_i&=exp(-(\phi d_{ij})^k),
    \end{aligned}
    \label{Geo_model}
\end{align}
where $d_{ij}$ is the distance between points $i$ and $j,$ $\phi$ controls the decay rate and $k$ is the smoothing parameter. The decay rate is modeled as $uniform(0.1,6)$. However, the parameter could be intuitively determined from the random process of interest \citep{Cramb2017}. Moreover, other valid spatial covariance function could replace the exponential function.

\textbf{--Stochastic Partial differential Equation (SPDE)}\\
SPDE, proposed by Lindgren \citep{lindgren2011explicit}, allows fitting a GRF with a continuously and smoothly decaying covariance function while gaining computation benefits from a GMRF representation. It represents a continuous random field using a discretely indexed spatial process. According to Blangiardo and Cemeletti \citep{Blangiardo2015}, the SPDE model is defined as follows. Let $\mathbf{z(s)}$ be as defined previously for a continuous spatial points $s \in D \subset {R}^d$. The SPDE model is given by
 \begin{align*}
     (k^2-\Delta)^{\alpha/2} \tau \mathbf{z(s)}=\field{W}(s),
 \end{align*}
$\Delta=\frac{\partial }{\partial^2s}$ is the Laplacian, $\alpha$ controls the smoothness,  $\tau$ controls the variance. $\field{W}(s)$ assumes a Gaussian spatial white noise process. However, $\field{W}(s)$ can also assume a Laplace noise, especially, when the data exhibit spikes and heavy tail. The solution to the above differential equation is a stationary Gaussian random field $\mathbf{z(s)}$ with Matern covariance structure $\sigma p(s_i,s_j)$, where $p(s_i,s_j)$ is a Matern correlation fuction between site $i$ and $j$ as defined previously,
\begin{align*}
    \alpha=v+\frac{d}{2},  \\
    \sigma^2=\frac{\Gamma(v)}{\Gamma(\alpha)(4\pi)^d/2 k^{2v})\tau^2},
\end{align*}
where $k=\frac{S_v}{r}$, $S_v$, $r$ and $v$ are as previously defined. The solution to the SPDE model is approximated by a basis function representation defined on a triangulation of the domain $D$. That is $z(s)=\sum_{g=1}^{G}\varphi_g(s)\Tilde{x}_g,$
G is the total number of vertices of the triangulation, $\{\varphi\}$ is the set of basis function, and $\{\Tilde{x}_g\}$ are zero-mean Gaussian distributed weights.\\

\subsection*{Class of Non-Gaussian random Fields Models}
A non-Gaussian process has increasingly been useful for modeling extreme ill-behaved random processes, such as predicting an earthquake and atmospheric temperature. This section discussed the asymmetric Laplace process,  inverse Wishart distribution, Pareto process, and Dirichlet process.

\textbf{--Asymmetric Laplace Process}\\
Suppose $Y(s)$ be an Asymmetric Laplace random vector with parameter $p$ and $\tau$,  $AL (p,0,\tau)$. According to Kuzobowski and Pogorski \cite{Kozubowski2000} and Fontanella  $et~ al$ \cite{,Fontanella2015} $Y(s)$ can be written as a sum of normal and exponential process, given by,
\begin{align}
    \begin{aligned}
    Y_p(\mathbf{s})&=\sqrt{\frac{2\xi(\mathbf{s})}{\tau p(1-p)}}Z(\mathbf{s})+\frac{1-2p}{p(1-p)}\xi(\mathbf{s}),\\
    Z(\mathbf{s})&\sim GP(\boldsymbol{0},\rho_z(\mathbf{s,s^*};\boldsymbol{\theta})),\\
    \xi(\mathbf{s})&\sim Gamma(1,\tau),
    \end{aligned}
    \label{asym_laplace}
\end{align}
where $\rho_z(\mathbf{s,s^*};\boldsymbol{\theta})$ is a valid spatial covariance function. $Z(\mathbf{s})$ is a Gaussian random field to accounts for spatial errors and it exist as a standard normal in its marginal form. $\xi(\mathbf{s})$ is marginally exponential distribution with rate $  \tau$, and it is conditionally independent of $Z(\mathbf{s})$. 
Thus, the conditional distribution of $Y_p(s)$ given $\xi(\mathbf{s})$ is given by,
\begin{align}
    Y_p(\mathbf{s})| \xi(\mathbf{s}) \sim N \Bigg (\frac{1-2p}{p(1-p)}\xi(\mathbf{s}),\frac{2}{\tau p(1-p) }\xi(\mathbf{s}) \Bigg ).
    \label{asym_laplace2}
\end{align}

Kristian and Alan \citep{Lum2012} discussed approaches to model $\xi(\mathbf{s})$ in a generalized quantile regression. In the spatial case, they defined $\xi(\mathbf{s})$ through CDF or copula transformation by letting $\xi(\mathbf{s})=-\frac{log(\Phi(V_{\xi}(\mathbf{s})))}{\tau}=F_{\tau}^{-1}(\Phi(V_{\xi}(\mathbf{s})))$, where $V_{\xi}(\mathbf{s})$ is  again a Gaussian process with a valid spatial covariance, $\Phi$ is a standard normal CDF, and $F_{\tau}$ is an exponential distribution CDF.

\textbf{--Multivariate Log-Gamma process}\\
Let matrix $\boldsymbol{V} \in \field{R}^n \times \field{R}^n$, and $\boldsymbol{\mathbf{\mu}} \in \field{R}^n$. Let $\mathbf{\gamma}=(\gamma_1,...,\gamma_n)^{'}$ be an $n$ mutually independent log-Gamma random variables with corresponding shape and scale parameter $\mathbf{\kappa}=(\kappa_1,...,\kappa_n)^{'}$ and $\mathbf{\alpha}=(\alpha_1,...,\alpha_n)^{'}$ respectively. That is $\gamma_i \sim LG(\kappa_i,\alpha_i)$. According to Bradley $et\; al.$ \citep{Bradley2018}, a $n$ dimensional vector $\mathbf{q}=\mu+V\gamma $ is a multivariate log-Gamma random variable and its distribution, denoted by  $MLG (\mathbf{\mu,V,\kappa,\alpha})$, mean and variance are given by, 
\begin{align}
\begin{aligned}
    f(q|\mu,V,\kappa,\alpha)&=\frac{1}{\big|\mathbf{VV}^{'}\big|^{1/2}}\Bigg ( \prod_{i=1}^{n}\frac{\alpha_i^{\kappa_i}}{\Gamma(\kappa_i)} \Bigg ) \exp \Bigg [ \mathbf{\kappa}^{'} \mathbf{V}^{-1}(\mathbf{q-\mu})-\mathbf{\alpha}^{'}\exp \{\mathbf{V^{-1}(\mathbf{q-\mu})\}} \Bigg ];\;\; \mathbf{q} \in \field{R}^n,\\
    \field{E}(\mathbf{q})&=\mathbf{\mu}+\mathbf{V}(\omega_0(\mathbf{\kappa})-log(\mathbf{\alpha})),\\
    Cov(\mathbf{q})&= \mathbf{V}\;diag(\omega_1(\mathbf{\kappa}))\;\mathbf{V^{'}},
    \end{aligned}
    \label{Mul_logGamma}
\end{align}
where $|\mathbf{A}|$ is a determinant of a matrix $\mathbf{A}$.\\
Consider a spatial random process  $Z(s)$ distributed as a multivariate log-Gamma, Yang $et\; al.$ \citep{Hu2018}, and Hu and Bradley \citep{Yang2019} described $Z(s)|\mathbf{\theta},\mathbf{\kappa},\mathbf{\alpha} \sim MLG (\mathbf{\mu,} \boldsymbol{\Sigma ^{1/2}},\kappa \mathbf{\field{I}},\alpha \mathbf{\field{I}})$, where $\Sigma=\sigma^2\rho(\mathbf{\theta},d(s,s^{*}))$ is a valid spatial covariance matrix, $\rho(\mathbf{\theta},d(s,s^{*}))$ is a correlation function, $d(s,s^{*})$ is a function of location $s$ and $s^{*}$, $\mathbf{\theta}$ is a vector of some parameters of interest, and $\field{I}$ is an identity matrix of appropriate dimension. An appropriate prior distribution are assigned to $\mathbf{\alpha}$ and $\mathbf{\kappa}$.\\

\textbf{--Student-t Process}\\
Some random processes exhibit heavy tail property, which may be inappropriate for using a Gaussian distribution, hence applying Student-t distribution in modeling spatial random process. Suppose $Z(s)$, as defined previously, be a random process that exhibits heavy tail property, observed at location $s\in \mathbf{D}$. This can be represented by setting $Z(s)$ to be distributed as a multivariate Student-t distribution $\mathbf{Z(s)} \sim t_n(0, {\Sigma},\mathbf{v}),$ where $\Sigma$ is a covariance matrix which is determined by a valid covariance function. The $i,jth$ element of $\Sigma$ is given by $Cov(Z(s_i),Z(s_j))$, and $\mathbf{v}$ is the degree of freedom. The $\Sigma$ is used to account for spatial dependencies.
Moreover, the covariance matrix can further account for spatially structured and unstructured (nugget) effects \citep{Schemmer2007}. The main challenge of the Student-t process is the difficulty of assigning appropriate prior specification on $\mathbf{v}$ to make inferences in a Bayesian analysis. Fonseca \citep{Fonseca2014} derived a Jeffreys-rule before $\mathbf{v}$, which lead to a proper posterior distribution. Moreover, Ordo{\~n}ez \citep{ordonez2020} extended it to a spatial domain and derived a reference prior to the joint spatial hyper-parameter and the degrees of freedom $v$.

\textbf{--Spatial Mixture model}\\
Green and Richardson \citep{Green2002} broaden the application of hidden Markov models for the random component $z_i$ by extending it to a spatial domain. It utilizes the model benefits of the Hidden Pott model. The allocation of the mixture components to clusters utilizes a spatial dependence structure, and the numbers of clusters are considered to be random. According to  Best and Thomson \citep{Best2005} the model specification is given in (\ref{Mix_model}). Areas estimated to belong in the same clusters need not be contiguous,
\begin{align}
    \begin{aligned}
    z_i&=log(\eta_{w_i)}, i=1,2,...,n,\\
    w_i&\in \{1,2,...,k\}\\
    p(\mathbf{w}|\boldsymbol{\psi},k)&=exp(\boldsymbol{\psi} U(\mathbf{w})-\delta_\mathbf{k}(\boldsymbol{\psi}))\\
    \eta_j &\sim Gamma (\alpha,\beta), j=1,2,...,k,\\
    k&\sim Uniform(1,c_{max}),
    \end{aligned}
    \label{Mix_model}
\end{align}
where $\boldsymbol{\psi}>0$, $c_{max}$ is the upper bound of the number of clusters, $U(\mathbf{w})=\sum_{i\sim i^{'}}{}\field{I}[z_i=z_{i^{'}}]$ is the number of same labels pairs to $i$ in a neighbouring area $i^{'}\neq i$ , parameter $\eta_j$ is associated with each component, and  $\delta_\mathbf{w}(\psi)=log(\sum_{w\in\{1,2,...,k\}^n}^{}\exp^{\psi U(\gamma)})$ is the normalizing constant, where the sum is the total possible ways for the allocation for the $n$ areas. The estimation of the numbers of clusters or components are obtained through reversible jump Markov chain Monte Carlo. Notice that $ p(\mathbf{w}|\psi,k)$ represents the probability function of the Pott model \citep{Li2019}.\\

\textbf{--Spatial Partition  model}\\
Leonhard and Ra{\ss}er \citep{knorr2000bayesian} proposed an approach for cluster detection, implemented using reversible jump Markov chain Morte Carlo. The location of clusters, the number of clusters, and the random process are unknown.  Best and Thomson \citep{Best2005} summarized the model as follows. A random number $k$ areas are selected as clusters, $g_j,j=1,2,...,k$. Conditioning on these $k$ areas, the remaining areas are assigned to their closest clusters $j\in \{1,2,...,k\}$. For a given cluster $k$, the positioning of clusters is assumed to have equally probable. All areas are assumed to be contiguous, unlike the mixture model. The model specification is given by, 
\begin{align}
    \begin{aligned}
     z_i&=\eta_{\gamma_i}, i=1,2,...,n,\\
    \gamma_i&\in \{1,2,...,k\}\\
   log( \eta_j) &\sim N(\alpha,\sigma^2), j=1,2,...,k,\\
    k&\sim Uniform(1,c_{max}) ~~or~Geometry,
    \end{aligned}
    \label{Spa_part_model}
\end{align}
where parameter $\eta_j$ is associated with each distinct geographic clusters.

\textbf{--Global Spline Mode}\\
Similarly to spartial models for area data, the spatial spline model assumes that each random field is centered on the centroid of a specific area in the spatial domain $D$. With the aim of separating a global geographical trend and local spatial trends, Lee and Durb{\'a}n \cite{lee2009smooth} proposed a two-dimensional P-spline at the centroids of each area in $D$ and was further extended to incorporate a random effect which is modeled with a CAR model. It was termed smooth-CAR models. According to Lee and Durb{\'a}n, the spatial P-spline model is described as follows. Suppose that $(s_{1i},s_{2j},z_{ij})$ are normally distributed spatial data, where $s_{1i},s_{2j}$ are respectively the longitude and latitude of the $ith$ centriod, and $z_{ij}$ is the response variable. The spline model is define as
\begin{align*}
    \bb{z}=f(s_{1},s_{2})+\m{\epsilon}=\m{B}\bb{\theta}+\m{\epsilon},
\end{align*}
 where $\bb{\theta}$ is a vector of coefficients , $\m{B}$ is a regression basis constructed from the coordinates $s_{1}$ and $s_{2}$, and $\m{\epsilon}\sim N(0,\sigma^2\mathbb{I})$. Then, P-spline approach penalizes the squared error loss with a penalty matrix $P$ depending on $\lambda$. That is,
\begin{align*}
    S(\bb{\theta};z,\lambda)=(z-\m{B}\bb{\theta})^{'}(z-\m{B}\bb{\theta})+\bb{\theta}^{'}\m{P}\bb{\theta}.
\end{align*}

When observed data violates the normality assumption and generally belongs to an exponential family with link function $g$. $\eta=g(\mu)=\m{B}\bb{\theta}$ with penalized sum of squares $l_{p}\bb{(\theta)}=l(\bb{\theta})+\bb{\theta}^{'}P{\theta}$ where $l(\bb{\m{\theta}})$ is the data likelihood. The penalized score is given as 
\begin{align*}
 (\m{B}^{'}\Tilde{\m{W}}_{\delta}\m{B}+\m{P})\hat{\theta}=(\m{B}^{'}\Tilde{\m{W}}_{\delta}\m{B}\Tilde{\m{\theta}}+\m{B}^{'}(y-\Tilde{\mu}),
\end{align*}
 where $\m{W}_{\delta}$ is a diagonal matrix with element $w_{ii}=\Big (\frac{\partial_{\eta_i}}{\partial \mu_i} \Big )^2 var(y_i)$. Where terms with tilde represents an approximate solution and terms with hat are the improved approximation.

\textbf{--Dirichlet Process (DP)}\\
Dirichlet process is a random process with sample functions almost surly a probability measure proposed by Ferguson \citep{Ferguson1973}. In the Bayesian framework, DP is a technique of analyzing non-parametric problems \citep{Antoniak1974}.
Let $\{Z(\mathbf{s}):\mathbf{s}\in \mathbf{D}\}, \mathbf{D}\subset{\field{R}^d}$ be the realizations with replicates of a spatial random field at distinct locations $\mathbf{s}$. Gelfand  $et\; al.$ \citep{Gelfand2005} described the spatial Dirichlet process as follows. Let $Z_t=(Z_t(s_1),...,Z_t(s_n))^{'}$, where $t$ represents replicates at site $s_i$. A Dirichlet process is a random probability measure defined on a measurable space $(\Omega,\mathbb{B})$, denoted by $DP(vG_0)$, where $v>0$ is the precision parameter and $G_o$ is a specific base probability distribution over $(\Omega,\mathbb{B}).$ Let $k_i, i=1,2,...$ be independent and identically distributed $Beta (1,v)$. The resulting random process for $Z(\mathbf{s})$ from Dirichlet process $DP(vG_0)$ defined  on  $(\Omega,\mathbb{B})$ can be written as $\sum_{i=1}^{\infty}\lambda_i\delta_{\theta_{i,D}}$, where $\delta_{k}$
represent a point mass at $k$ and $\lambda_1=k_1$,  $\lambda_i=k_i\prod_{j=1}^{i-1}(1-k_j),i=2,3,...$ , and $\{\theta_i(s):s \in D\}$ are realizations from base probability $G_0$ which can possibly be a stationary Gaussian process. Notice that the DP process are independent, however, MacEachern \citep{MacEachern2000} relaxed this condition by deriving the dependent $DP$. Moreover, Gelfand $et\;al.$ \citep{Gelfand2005} extended the dependent $DP$ to account for spatial dependence. The property that DP process is almost surely discrete restricts its applications to a wide class of continuous problems. Hence, Antoniak \citep{Antoniak1974} derived a mixtures of DP to circumvent the problems, and thus extended it to handle continuous cases.
\subsection*{Statistical Models}
\textbf{--Generalized Linear Mixed Model}\\
 A classical linear model formulation is given by
\begin{align}
\begin{aligned}
y_i&\sim N(\mu_i,\sigma^2)\\
y_i&=\beta_0+\sum_{j=1}^{J}\beta_ix_{1i}+\epsilon_i \\
  \epsilon_i&\sim N(0,\sigma^2),~~i=1,2,...n\\
  \field{E}(y_i |\mu_i)&=\mu_i\\&=\beta_0+\sum_{j=1}^{J}\beta_ix_{1i},
  \end{aligned}
    \label{reg_modem}
\end{align}
 and the matrix form is given by,
\begin{align}
 \boldsymbol{Y}&=\boldsymbol{X}\mathbf{\beta}+\epsilon,
 \label{reg_matrix}
\end{align}
where the latent field $\beta=(\beta_0,\beta_1,...,\beta_p)^{'}$ defines the relationship between response variable $\boldsymbol{Y}$ and covariate $\boldsymbol{X}.$ \citep{dey2000}.
Each outcome $y_i$ is assumed to be generated according to a Gaussian distribution. The mean depends on related covariates through $\field{E}(y|\mu).$ A generalized form of (\ref{reg_matrix}) relaxes the assumption that the errors are Gaussian distributed and each outcome is generated from a non-Gaussian distribution such as Binomial, Poisson, Beta, among others \citep{fahrmeir2013}. It allows these random processes to be modeled through a link function and allows the magnitude of the measurement error to be a function of the predicted estimates. That is,
\begin{align}
\begin{aligned}
y_i&\sim \mathbb{P}(.|\theta),\\
g(\theta)&=\beta_0+\sum_{j=1}^{J}\beta_jx_{ji}+\epsilon_i,
\end{aligned}
\label{GLM}
\end{align}
where $g$ is an appropriate link function such as the $logit$ for a Binomial model and $log_e$ for the Poisson model. The mixed form of  Equation (\ref{GLM}) incorporates a function $f(\dot)$ to relax the linearity assumption on covariates or to introduce a random effect usually modeled as a random walk,  auto-regressive, or penalized spline models \citep{fahrmeir2013}. The mixed form is given in (\ref{GLMM})
\begin{align}
\begin{aligned}
y_i&\sim \mathbb{P}(.|\theta)\\
g(\theta_i)=&\beta_0+\sum_{j=1}^{J}\beta_jx_{ji}+\sum_{k=1}^{K}f_k(z_{ki})+f^{*}(v_i)+\epsilon_i,
\end{aligned}
\label{GLMM}
\end{align}
where $z_k$ is the $kth$ random effect and $v_i$ is a spatial effect assigned spatial prior distribution. The latent field of interest is given as $\Theta=\{\beta,f,v\}.$ Equation \ref{GLMM} is termed the Generalized Linear Mixed Model (GLMM).  In a Bayesian setting, all the parameters in the model are considered random, and appropriate prior distributions are assigned. In the absence of prior information, a non-informative prior is assigned to $\beta \sim Normal (0,10^6)$ and $log(1/\sigma^2)\sim log gamma(1,10^{-5})$ \citep{Blangiardo2015}. A Bayesian hierarchical model contains the data, prior, and hyperprior stages,
\begin{align}
\begin{aligned}
    y_i|\Theta,\phi &\sim \mathbb{P}(y|\Theta,\phi),\\
    \Theta|\phi &\sim \mathbb{P}(\Theta|\phi),\\
    \phi &\sim \mathbb{P}(\phi),
\end{aligned}
\label{HGLMM}
\end{align}
where $  \theta_i\in \Theta|y$ is of main interest.
These models are usually referred to as latent Gaussian models, which are flexible and can accommodate a wide range of models.

\textbf{--Survival Model}\\
A survival data analysis models the time to the event occurrence. This can be time to death of subject under study, process failure time, or time to radioactive emission. Let $T_i$ be a random variable of survival times, then $S(t|\tau)=\mathbb{P}(T>t|\tau)$
is the survival function that, for example, determines the probability that a patient survives over time $t$. It is assumed that all subject are alive, $S(0|\tau)=1$, and all subject will eventually die $\lim_{t\to +\infty}S(t|\tau)=0$. The survival function is expressed as a distribution function $F(t)=\mathbb{P}(T<t|\tau)=1-S(t|\tau)$ with probability distribution $f(t|\tau)$. The hazard function $h(t)$ measures the probability that an event will occur at a small instance $\Delta t$ after the subject has survived through time $t$. That is,
    \begin{align*}
        h(t)&=\lim_{\Delta t\to 0}\frac{\mathbb{P}(T<t+\Delta t|T>t)}{\Delta t},\\
        H(t)&=\int_{0}^{t}h(u)du,
    \end{align*}
where $H(t)$ is the cumulative harzard function. 

Let $(t_i,\delta_i,\boldsymbol{x_i})$ represents a survival data, where $T_i$ represents survival times of the $ith$ subject of interest, $\boldsymbol{x_i}$ is a predictor variables, and $\delta_i=\field{I}(T_i\leq C)$, where $C$ is a censoring threshold set a priori. Considering the covariates, the likelihood of the model is expressed by,
\begin{align}
\begin{aligned}
  g(\tau_i)&=\beta_0+\sum_{j=1}^{J}\beta_jx_{ji}+\sum_{k=1}^{K}f(z_k)+v_i+ \epsilon_i,\\
    L(\theta|t)&=\prod_{i=1}^{n}f(t_i|\tau)^{\delta_i}S(t_i|\tau)^{1-\delta_i},
\end{aligned}
\label{Survival}
\end{align}
where $\theta= \{\beta_1,...,\beta_j,\tau,\Phi \}$, $\beta_i,$ is a fixed effect, $z_k$ is a random effect, $v_i$ with parameter $\Phi,$ is a spatial random effect assigned a spatial prior in a Bayesian framework \citep{dasgupta2014,onicescu2018}. $ f(t_i|\tau) $ can assume Exponential, Weibull, Logistics, or lognormal distribution to list a few. The censoring status $\delta$  controls the contribution to the likelihood of subjects that experienced the event and those that survived through the entire study period. Kaplan-Meier provides non-parametric estimates of the survival curves. The proportional hazards model and accelerated failure time model are the most frequently used frequentist parametric methods to analyze survival data \cite{kwak2017central}.

\textbf{--Bayesian Spatial Econometrics}\\
Spatial analytical tools are widely used in Economics to quantify the spatial dependencies and heterogeneity of economic variables. Referred to as spatial econometrics, it extends the traditional econometrics to a spatial domain. The most frequently used models are the Spatial Autoregressive (SAR), Spatial Error Model (SEM), and Spatial Durbin Model (SDM). The models are briefly described as follows.\\
Let $Y$ be a response variable assuming a linear relationship with explanatory variables $\boldsymbol{X}$, the SAR model is represented by,
\begin{align}
    \begin{aligned}
        \boldsymbol{Y}&=\rho \boldsymbol{W}Y+\boldsymbol{X}\boldsymbol{\beta}+\epsilon,\\
        \epsilon &\sim MVN(0,\sigma^2\mathbb{I}).
    \end{aligned}
    \label{SAR}
\end{align}
$\boldsymbol{W}$ be a spatial weight matrix, $\rho$ is a spatial autocorrelation parameter, and $\boldsymbol{\beta}$ is a vector of the regression parameters. The SAR model assumes that the dependent variable is spatially autocorrelated \citep{jensen2013}. In contrast, the SEM model assumes the error is a spatial correlation. The SEM formulation is given by,
\begin{align}
    \begin{aligned}
        \boldsymbol{Y}&=\boldsymbol{X}\boldsymbol{\beta}+u,\\
        u&=\rho \boldsymbol{W}u+\epsilon,\\
        \epsilon &\sim MVN(0,\sigma^2\mathbb{I}).
    \end{aligned}
    \label{SEM}
\end{align}
The SDM is an extension of the SAR model which assumes that the dependent variable and covariates are spatially correlated. The formulation is given by,
\begin{align}
    \begin{aligned}
        \boldsymbol{Y}&=\rho \boldsymbol{W}Y+\boldsymbol{W}\boldsymbol{X}\boldsymbol{\beta}+\epsilon,\\
        \epsilon &\sim MVN(0,\sigma^2\mathbb{I}).
    \end{aligned}
    \label{SDM}
\end{align}

According to LeSage and Pace \cite{lesage2008introduction}, SDM is appropriate when the included covariates are correlated with spatially correlated variables not included in the model. In the next sections, we present each class's main idea of the spatial literature's estimation methods.
\subsection*{Estimation Method}
\textbf{--Maximum Likelihood Estimation}\\
The maximum likelihood estimation approach involves an optimization problem to determine the best sets of distribution parameters representing data. In spatial statistics, the MLE method is usually used to estimate global spatial dependencies in a spatial econometric model. Authors often compare its estimates with one obtained in a Bayesian inferential framework and highlight the MLE approach's inadequacies to spatial statistical inferences \citep{neill2006,pilz2012,porto2013,yu2008}. 

Let $y_i\sim f(.|\bb{\theta_i})$, where $\bb{\theta_i\in \Theta}$ is a parameter of interest and $\bb{\Theta}$ is the parameter space. $y_i$ need not be independent and identical. The estimation procedure involves computing the likelihood (log-likelihood) of the data distribution and determine the sets of parameter $\theta^{*}=\{\theta_i^{*}: \theta_i^{*}\in \bb{\Theta},i=1,2,...,n\}$ where the likelihood is maximum. For independent $y_i$, $i=1,2,...,n$, the likelihood is given by,
\begin{align}
\begin{aligned}
    L(\bb{\theta}|\m{y})=\prod_{i=1}^{n} f(y_i|\bb{\theta_i}),\\
    \bb{\theta}^{*}=\underset{\bb{\theta}}{\arg\max} ~~log \big( L(\bb{\theta}|\m{y})\big ).
\end{aligned}
    \label{MLE1}
\end{align}

An equivalent approach to the classical MLE when the dimension of the model parameters is large or complex for estimation is the quasi-Maximum likelihood estimation (QMLE). Unlike the MLE, the QMLE maximizes a function $log\; L^{*}(\bb{\theta}|\m{y})$ that is related to the logarithm of the likelihood function $ L(\bb{\theta}|\m{y})$. Su and Yang \citep{su2015} proposed a QMLE of dynamic panel models with spatial errors and  was further broadened by Yu, De Jong, and Lee $et.al$ \citep{yu2008}. An equivalent approach to MLE in the Bayesian setting is the Maximum a Posteriori (MAP). It involves maximizing the posterior conditional probability,
\begin{align}
   \bb{\theta}^{*}= \underset{\bb{\theta}}{\arg\max}\;p(\bb{\theta})p(\m{y}|\bb{\theta}),
\end{align}
where $p(\bb{\theta})$ is the prior distribution and $p(\m{y}|\bb{\theta})$ is the data likelihood defined as $L(\m{y}|\bb{\theta})$ in (\ref{MLE1}).

\textbf{--Expectation Maximization (EM Algorithm)}\\
The maximum likelihood estimation limitation is the assumption that the variables that generate the process are all observable. In practice, this assumption rarely holds. One possible choice to overcome the limitation is the estimation through the EM algorithm. 

The EM algorithm, proposed by Dempster $et~al.$ \citep{dempster1977}, fits a model to a latent representation of a data rather than merely fitting distribution models. It can work well in data that contains unobserved (latent) variables. The algorithm, an iterative method, has two major stages: estimating the latent variables (E-step) and maximizing the model parameters given the data and the estimated variables (M-step). 

Let $\boldsymbol{\theta}$ be initialized model parameter, the E-step is used to update the latent space variables $z$, usually discrete or cluster in particular, through  $p(\mathbf{z}|\mathbf{y},\boldsymbol{\theta})$, where $\mathbf{y}$ is the observed data. To update $\boldsymbol{\theta}$, the expectation $\field{E}_{\mathbf{z|y,}\boldsymbol{\theta^{*}}}log(p(\boldsymbol{y|z},\boldsymbol{\theta}))$ is computed. $\boldsymbol{\theta^*}$ represents the previous parameter and $\boldsymbol{\theta}$ is the potential new parameter of the model 
\begin{align}
 \field{E}_{\mathbf{z|y,}\boldsymbol{\theta^{*}}}log(p(\boldsymbol{y|z},\boldsymbol{\theta}))=\sum_{i=1}^{n}\sum_{j=1}^{k} p(z_i=j|\mathbf{y},\boldsymbol{\theta^{*}})log[p(z_i=j|\boldsymbol{\theta})p(y_i|z_i=j,\boldsymbol{\theta})].
\end{align}
In the M-step, the EM-algorithm maximizes the model parameter in the equation
\begin{align}
    \boldsymbol{\theta^{*}}=\underset{\theta}{\arg\max}~~ \field{E}_{\mathbf{z|y,}\boldsymbol{\theta^{*}}}log(p(\boldsymbol{y|z},\boldsymbol{\theta})).
\end{align}

The iteration continues until the difference between the current and the previous expectation is lesser than $\epsilon>0$ set at the initial stage.

\textbf{--Markov Chain Monte Carlo (MCMC)}\\
Given a posterior distribution 
\begin{align}
    p(\boldsymbol{\theta,\psi}|\mathbf{y})\propto p(y|\boldsymbol{\theta,\psi})\times p(\boldsymbol{\theta,\psi}),
    \label{MC}
\end{align}
and assuming that the posterior $p(\theta|y,\Psi)$ is of known form, such as a standard probability distribution, we can resort to Monte Carlo approach to approximate posterior quantities $h(\bb{\theta})$,
\begin{align*}
    \mathbb{E}(h(\bb{\theta})|\m{y})=\int_{\bb{ \theta}\in\Theta}p(\theta|\m{y})d\theta=\int_{\bb{\theta}\in\bb{\Theta}} \int_{\bb{\psi}\in \bb{\Psi}} h(\theta)p(\theta|\bb{\psi},y)p(\bb{\psi}|\m{y})d\bb{\psi} d\bb{\theta},
\end{align*}
which could be  the mean, median or  higher moments. The procedure consist of simulating an $m$ random samples from $p(\theta|y)$, say $\{\theta^{1},\theta^{2},...,\theta^{m}\}$ and evaluate  the unknown quantity $h(\theta)$ using the empirical average 
\begin{align}
\mathbb{E}(\hat{h(\theta)}|\mathbf{ y})=\frac{\sum_{i=1}^{m}h(\theta^{i})}{m}.
\label{Emperical}
\end{align}

Under the Law of Large Numbers, the empirical distribution will converge to the true distribution. In the case of a joint posterior distribution, an approximation of the posterior marginals are achieved by first sampling from a marginal distribution of a subset of parameters given its complements and then use it to evaluate the full joint distribution.\\

In practice, the posterior distribution's functional form is unknown or complex, and independent samples are not feasible. An alternative approach comprises generating independent samples from an importance distribution $\mathbb{q}(\theta|y)$ which is a close distribution to $\mathbb{p}(\theta|y)$. Empirically, the quantity $h(\theta)$ is obtained as  
\begin{align}
    \mathbb{E}(\hat{h(\theta)})=\sum_{i=1}^{m}\frac{h(\theta_i)\mathbb{p}(\theta_i)}{\mathbb{q}(\theta_i)m}.
\end{align}
This approach is not trivial for a large number of dimensions of $\theta$ \citep{hastings1970}. A more widely used alternative approach comprises generating correlated samples by running a Markov chain whose stationary distribution converges to the posterior density. The posterior summaries are computed from these samples using the empirical method, as described above. Suppose $\chi$ is the state space of the posterior distribution.  As stated by Blangiardo \citep{Blangiardo2015}, the convergence of the Markov chains stationary distribution to the posterior distribution requires that the Markov chains are irreducible (the chain has a positive probability of reaching all region of $\chi$ regardless of the starting point), recurrency (the limit of the probability of the chain visiting a subset $\chi$ infinitely many times is 1), and aperiodic (the chain do not circles when exploring $\chi$). The highlighted procedure is referred to as MCMC. Gibbs sampler and Metropolis-Hastings algorithm are the most frequently used standard MCMC algorithm in Bayesian inference literature. For a description of these algorithms, see \citep{Blangiardo2015} (pp. 91-103).

\textbf{--Integrated Nexted Laplace Approximation (INLA)}\\
INLA, proposed by Rue $et~al$ \citep{rue2009}, is an alternative approach to the estimation of posterior marginals. It has gained considerable attention and has been proven to outperform the MCMC approach in computational speed  \citep{rue2009} . The availability of the  $R-INLA$ simplifies the implementation of the approach which authors from the diverse field have found useful and easy.

Again, consider the posterior distribution 
\begin{align}
    p(\theta_i|\m{y}) \int p(\theta_i,\bb{\psi}|\m{y})d\bb{\psi}= \int  p(\theta_i|\boldsymbol{\psi},\m{y})p(\bb{\psi}|\m{y})d\bb{\psi}.
    \label{inla1}
\end{align}

The objective is to obtain the posterior marginals $p(\theta_i|y)$ for each parameter in the vector and the estimates of the hyperparameters given by,
\begin{align}
    p(\boldsymbol{\psi}_k|\mathbf{y}) \int p(\boldsymbol{\psi}_k|\mathbf{y})d\boldsymbol{\psi_{-k}}.
    \label{inla2}
\end{align}

The INLA approach utilizes the model assumptions to approximate the marginal posterior distribution and its moments based on Laplace approximation  \citep{tierney1986}. According to  \citep{Blangiardo2015,blangiardo2013} INLA approximation follows the following steps. Firstly, the posterior marginals of the hyperparameters are approximated, that is,
\begin{align*}
    p(\boldsymbol{\psi}|\mathbf{y})&=\frac{p(\boldsymbol{\theta},\boldsymbol{\psi}|\mathbf{y})}{p(\boldsymbol{\theta}|\boldsymbol{\psi},\mathbf{y})}\propto \frac{p(\boldsymbol{\psi})p(\boldsymbol{\theta}|\boldsymbol{\psi})p(\mathbf{y}|\boldsymbol{\theta})}{p(\boldsymbol{\theta}|\boldsymbol{\psi},y)},\\
    &\approx  \at {\frac{p(\boldsymbol{\psi})p(\boldsymbol{\theta}|\boldsymbol{\psi})p(\mathbf{y}|\boldsymbol{\theta})}{\Tilde{p}(\boldsymbol{\theta}|\boldsymbol{\psi},\mathbf{y})}}{\boldsymbol{\theta}=\boldsymbol{\theta^{*}}(\boldsymbol{\psi})} \vcentcolon \Tilde{p}(\boldsymbol{\psi}|\mathbf{y}),
\end{align*} 
where $\Tilde{p}(\boldsymbol{\theta}|y)$ is a Gaussian approximation for $p(\boldsymbol{\theta}|y)$ and $\theta^{*}$ is the mode. Secondly, the parameter vector is partitioned such that $\boldsymbol{\theta}=(\theta_i,\theta_{-i})$ and again approximated using Laplace procedure to obtain
\begin{align*}
    p(\theta_i|\boldsymbol{\psi},\mathbf{y})=\frac{p(\theta_i,\boldsymbol{\theta_{-i}}|\boldsymbol{\psi},\mathbf{y})}{p(\boldsymbol{\theta}_{-i}|\boldsymbol{\theta_i},\boldsymbol{\psi},\mathbf{y})}\approx \at {\frac{p(\boldsymbol{\theta},\boldsymbol{\psi}|\mathbf{y})}{\Tilde{p}(\boldsymbol{\theta}_{-1}|\theta_i,\boldsymbol{\psi},\mathbf{y})}}{\boldsymbol{\theta_{-i}}=\boldsymbol{\theta^{*}}_{-i}(\theta_i,\boldsymbol{\psi})} \vcentcolon \Tilde{p}(\theta_i|\boldsymbol{\psi},y).
\end{align*}

To bypass the computational complexity of computing $\Tilde{p}(\theta_i|\boldsymbol{\psi},y)$, INLA explores the marginal joint posterior for the hyperparameters $p(\boldsymbol{\psi}|\mathbf{y})$ in a grid search to select an important points $\{ \boldsymbol{\psi_k}\}$ jointly with a corresponding set of weights $\{ \Delta_k\} $ to give approximates to the posterior to the hyperparameters. Each marginals $\Tilde{p}(\boldsymbol{\psi_k}|\mathbf{y}) \forall k$ can be obtained using log-spline interpolation bases on selected $\boldsymbol{\psi_k}$ and $\Delta_k$. For each $k$ the conditional posterior $\Tilde{p}(\theta_i|\boldsymbol{\theta_i}|\psi,y)$ is computed and a numerical integration
\begin{align}
    \Tilde{p}(\boldsymbol{\theta_i}|\mathbf{y}) \approx \sum_{k=1}^{K}\Tilde{p}(\theta_i|\bb{\psi_k},\mathbf{y})\Tilde{p}(\boldsymbol{\psi_k}|\mathbf{y}).
    \label{inla3}
\end{align}
is then used to obtain $\Tilde{p}(\theta_i|\boldsymbol{\psi},y)$.
\end{document}